%% file: final.tex
\documentclass[final]{siamltex}

\usepackage{graphicx}
\usepackage{amsfonts}
\usepackage{amsmath}
\usepackage{amssymb}
\usepackage{bm}
\usepackage{amsbsy}
\usepackage{algorithmic}
\usepackage{algorithm}
\usepackage{subfigure}
\usepackage{listings}
\usepackage{tikz}
\usetikzlibrary{shapes}
\usetikzlibrary{plotmarks}
\usepackage{stackrel}
\usepackage{url}
\usepackage{xspace}

\usepackage[T1]{fontenc}
\usepackage[ansinew]{inputenc}
\usepackage{lmodern}

\usepackage[T1]{fontenc}
\usepackage[ansinew]{inputenc}
\usepackage{lmodern}

\usepackage{ifpdf}
\ifpdf
  \usepackage{epstopdf} 
\fi

\usepackage[letterpaper,centering]{geometry}
\usepackage[colorlinks=true, citecolor=blue, filecolor=black,linkcolor=red, urlcolor=blue, hypertexnames=false, pdftex]{hyperref}

\usepackage{todonotes}


\title{Efficient Localization of Discontinuities in Complex Computational Simulations}

\author{Alex Gorodetsky\footnotemark[1] \and Youssef Marzouk\footnotemark[2]}

\begin{document}

\maketitle

\renewcommand{\thefootnote}{\fnsymbol{footnote}}

\footnotetext[1]{Department of Aeronautics and Astronautics, Massachusetts Institute of Technology, Cambridge, MA
02139, USA, \texttt{goroda@mit.edu}.}

\footnotetext[2]{Department of Aeronautics and Astronautics, Massachusetts Institute of Technology, Cambridge, MA
02139, USA, \texttt{ymarz@mit.edu}.}

\renewcommand{\thefootnote}{\arabic{footnote}}

\begin{abstract} 
  Surrogate models for computational simulations are input-output
  approximations that allow computationally intensive analyses, such
  as uncertainty propagation and inference, to be performed
  efficiently. When a simulation output does not depend smoothly on
  its inputs, the error and convergence rate of many approximation
  methods deteriorate substantially. This paper details a method for
  efficiently localizing discontinuities in the input parameter
  domain, so that the model output can be approximated as a piecewise
  smooth function.
  The approach comprises an initialization phase, which uses
  polynomial annihilation to assign function values to different
  regions and thus seed an automated labeling procedure, followed by a
  refinement phase that adaptively updates a kernel support vector
  machine representation of the separating surface via active
  learning. The overall approach avoids structured grids and exploits
  any available simplicity in the geometry of the separating surface,
  thus reducing the number of model evaluations required to localize
  the discontinuity.
  The method is illustrated on examples of up to eleven dimensions,
  including algebraic models and ODE/PDE systems, and demonstrates
  improved scaling and efficiency over other discontinuity
  localization approaches.
\end{abstract}

\begin{keywords}
{discontinuity detection},
{polynomial annihilation},
{function approximation},
{support vector machines},
{active learning},
{uncertainty quantification}
\end{keywords}

\pagestyle{myheadings}
\thispagestyle{plain}
\markboth{\MakeUppercase{A.\ Gorodetsky and Y.\ M.\ Marzouk}}{\MakeUppercase{Efficient localization of discontinuities}}

\input{introduction}

\input{background}

\input{algorithm}
\input{results}

\input{conc-ack}

\bibliography{mybib}
\bibliographystyle{siam}

\end{document}

%% file: introduction.tex
\section{Introduction}

Many applications of uncertainty quantification, optimization, and control must invoke models accessible only through computational simulation. These tasks can be computationally prohibitive, requiring repeated simulations that may exceed available computational capacity. In these circumstances, it useful to construct surrogate models approximating the simulation output over a parameter domain of interest, using a limited set of simulation runs. The construction of surrogates is essentially a problem in function approximation, for which an enormous variety of approaches have been developed. One broad category of approximations involves parametric or semi-parametric representations---for instance, polynomial expansions obtained via interpolation, projection, or regression~\cite{Xiubook,Ghanembook,Xiu2005,Conrad2012,Eldred2008}. Another category involves nonparametric approximations such as Gaussian process regression~\cite{Rasmussen2006}, frequently used in the statistics community for the ``emulation'' of computer models \cite{Conti2010,SacksWelch1989} 

Almost all of these approximation methods deteriorate in efficiency when faced with discontinuities in the model output, or even its derivatives, over the range of input parameters. Yet discontinuities frequently arise in practice, e.g., when systems exhibit bifurcations with respect to uncertain input parameters or enter different regimes of operation depending on their inputs. Examples include ignition phenomena in combustion kinetics~\cite{Najm2009chem}, bifurcations in climate modeling~\cite{Webster2007}, switch-like behavior in gene expression~\cite{Gardner2000}, and in general, dynamical systems with multiple equilibria. In all of these applications, being able to \textit{localize} the discontinuity would enable significant efficiency gains in the construction of output surrogates. Moreover, localizing a discontinuity may be of standalone interest; since a discontinuous response may be a defining feature of the system, learning exactly which input or parameter regimes yield different behaviors can lead to a more fundamental understanding of the system's dynamics.

In this work, we will focus on piecewise smooth model responses; in other words, we assume that the parameter space contains one or more ``separating surfaces'' that bound regimes over which the model output is a smooth function of its parameters. Jumps in the model output occur across a separating surface. The separating surface may itself be relatively smooth and well approximated by techniques which take advantage of this regularity. The major contribution of this work is then an unstructured approach for identifying and refining a functional description of the separating surface. Our approach uses guided random sampling to place new model evaluation points in the vicinity of the discontinuity. These points are labeled and used to drive a kernel support vector machine classifier, which yields a nonparametric description of the discontinuity location. The entire approach is iterative: following an initialization and labeling phase, it employs a cycle of active learning, labeling, and classification. The overall algorithm uses significantly fewer model evaluations and exhibits improved scaling with parameter dimension compared to current discontinuity detection techniques. It contrasts with efforts that have generally attempted to create a dense and structured grid of model evaluations surrounding the separating surface.

The remainder of this paper is organized as follows. Section~\ref{sec:background} reviews current techniques for discontinuity detection and for approximating discontinuous model responses. Section~\ref{sec:tools} describes the algorithmic building blocks from which we construct our approach. In Section~\ref{sec:method} we detail the discontinuity detection algorithm itself. In Section~\ref{sec:results} we report on numerical experiments with this algorithm: discontinuity detection problems of increasing dimension, problems that vary the complexity of the separating surface, and several benchmark ODE and PDE problems drawn from the literature.

%% file: background.tex
\section{Background} \label{sec:background}
Approximation schemes for discontinuous model outputs typically attempt to transform the problem into one that can be tackled with classical approximation methods for smooth functions. These transformations can roughly be divided into three categories: local approximations, edge tracking, and global approximations. 

Local approximations may involve either decomposing the parameter space in a structured manner (e.g., into hypercubes) or utilizing local basis functions. Examples of parameter space decomposition include multi-element gPC \cite{Wan2005} or treed Gaussian processes
\cite{Gramacy2004, Gramacy2008, Bilionis2012}; examples of local basis functions include wavelets \cite{LeMatre2004, LeMatre2004b} or particular forms of basis enrichment~\cite{Ghosh2008}. These techniques attempt simultaneously to find the discontinuity and to build the approximation. Edge tracking techniques, on the other hand, separate discontinuity localization from approximation and concentrate on the former~\cite{Jakeman2011}. Another approach that separates discontinuity localization from approximation can be found in~\cite{SargsyanSafta2012}, where a Bayesian classification method is used to represent the separating surface, and the two resulting classes are mapped to distinct hypercubes wherein the functions are approximated by polynomial chaos expansions. Finally, there exist global methods that attempt to directly mitigate the Gibbs phenomena arising from approximating discontinuous functions with a smooth basis. One such effort~\cite{Chantrasmi2009} employs Pad\'{e}-Legendre approximations in combination with filtering to remove spurious oscillations. These techniques have been successfully demonstrated in low-dimensional parameter spaces. 

Below we elaborate on domain decomposition and edge tracking methods, as they provide useful inspiration for our present method.

\subsection{Domain decomposition and local approximation}
Decomposition techniques approach the approximation problem by breaking a domain containing a discontinuity into sub-domains containing smooth portions of the function of interest. Examples are given in \cite{Wan2005} and \cite{Archibald2009}.  These algorithms may be distinguished according to three attributes: refinement criteria, point selection scheme, and approximation type. Refinement criteria are indicators that specify the need for additional function evaluations; for example, they may be tied to an estimate of the discontinuity location, or to a local indicator of error in the function approximation. Point selection describes the manner in which additional function evaluations are added, e.g., deterministically or randomly, near or far from previous evaluations, etc. Finally, the approximation type may involve a choice between low- or high-order polynomials, parametric or nonparametric schemes, etc. These choices are closely intertwined because the refinement criteria and point selection scheme are often guided by the type of approximation performed in each subdomain.
Many current techniques for domain decomposition rely on adaptively partitioning the parameter domain into progressively smaller hypercubes. Building approximations on these Cartesian product domains is convenient, but can be computationally expensive, particularly when separating surfaces are not aligned with the coordinate axes. These difficulties are exacerbated as the parameter dimension increases.

Related to domain decomposition are approaches that use local basis functions to capture sharp variations in model output \cite{LeMatre2004, LeMatre2004b} . These approaches also tend to rely on the progressive refinement of hypercubes. Localized bases are extensively employed in the image processing community \cite{Kingsbury1999}; for example, images often have sharp edges that are accurately represented with wavelets \cite{Meyer1993, Daubechies1988} and other functions with local support. In practice, these basis functions are often deployed within an adaptive approach that yields a dense grid of function evaluations surrounding the discontinuity. The resulting model runs occur in similar locations and with a number/computational cost similar to domain decomposition methods.

\subsection{Edge tracking}
An alternative and rather efficient algorithm for discontinuity localization has been developed in~\cite{Jakeman2011}. As noted above, the algorithm focuses on searching for a discontinuity and developing a description of the separating surface, rather than on approximating the true model. In particular, the algorithm progressively adds points by ``walking'' along the discontinuity (i.e., edge tracking), while using polynomial annihilation (PA) along the coordinate axes as an indicator of the discontinuity's existence and location. This procedure uses an adaptive divide-and-conquer approach to initially locate the separating surface. After edge tracking is complete, new evaluation locations are classified---i.e., deemed to lie on one side of the separating surface or the other---using a nearest neighbor approach. The majority of the computational effort is thus spent evaluating the model near the separating surface, such that the resulting set of points becomes an evenly spaced grid surrounding it. Having located the discontinuity, function approximation can then proceed on each surrounding subdomain. For example, edge tracking is coupled with the method of least orthogonal interpolation~\cite{Narayan2012} in~\cite{Jakeman2013}.

Because a greater fraction of its computational effort is spent evaluating the model close to the separating surface, edge tracking is more efficient at discontinuity localization than the domain decomposition methods presented earlier. The method proposed in this paper capitalizes on this philosophy and aims for further improvement by taking advantage of the regularity of the separating surface. Rather than walking along the surface with steps of fixed resolution, we introduce a new method for \textit{sampling} in the vicinity of the discontinuity and for efficiently describing the geometry of the separating surface given an unstructured set of sample points. These developments will be detailed below.

%% file: algorithm.tex
\section{Algorithmic ingredients of our approach}\label{sec:tools}

The new discontinuity detection algorithm described in this paper is founded on several tools common in the machine learning and spectral methods communities. These tools will be used to address three problems arising in the approximation of high-dimensional discontinuous functions. The first problem involves identifying the separating surface and estimating the jump size of the discontinuity across it.  The jump size is a local measure of the difference between the function values on either side of the separating surface. We will solve this problem using \textit{polynomial annihilation}. The solution will also provide a method for labeling function evaluations on either side of the separating surface, based upon their function value.  The second problem is to find an efficient representation of the geometry of the separating surface; to this end, we will employ a nonparametric approximation using \textit{support vector machines} (SVM). The final problem involves determining locations at which to evaluate the function in order to best refine the approximation of the separating surface. Our solution to this problem will employ \textit{uncertainty sampling} techniques.

\subsection{Polynomial annihilation}\label{sec:pa}
Polynomial annihilation is used in order to measure the size of a discontinuity or region of rapid change in a function. This measurement is vital for determining the region to which new function evaluations belong. Following \cite{Archibald2005}, a description of one-dimensional polynomial annihilation is given here. The local size of the discontinuity is described in terms of the jump function evaluated at a particular location in the parameter space. Suppose that $x \in \mathbb{R}$ and $f: \mathbb{R}\rightarrow\mathbb{R}$. The jump function, $[f](x)$, is defined to be
\begin{equation}
    [f](x) = f(x+) - f(x-),
\end{equation}
where $ f(x-) = \lim_{\Delta \rightarrow 0} f(x-\Delta) $ and $f(x+) = \lim_{\Delta \rightarrow 0} f(x+\Delta).$
Therefore, $[f](x)$ is non-zero when the function is discontinuous at $x$, and it is zero otherwise. The main result of polynomial annihilation is the approximation $L_mf$ to the jump function. This approximation has the form 
\begin{equation}
L_mf(x) = \frac{1}{q_m(x)}\sum_{x^l \in \mathcal{S}(x)}c_l(x)f(x^l),
\label{eq:jumpapprox}
\end{equation}
where the set $\mathcal{S}(x)$ is a ``stencil'' of points $(x^l)$ around $x$. The coefficients $(c_l)$ are calculated by solving the system of equations
\begin{equation}
\sum_{x^l \in \mathcal{S}(x)} c_l(x)p_i(x^l) = p_i^{(m)}(x), \quad i = 0 \ldots m,
\label{eq:tres}
\end{equation}
where $m$ is the order of desired annihilation and $p_i$ comprise a basis for the space of univariate polynomials of degree less than or equal to $m$. An explicit expression for each $c_l$, derived in~\cite{Archibald2005}, is
\begin{equation}
c_l(x) = \frac{m!}{\prod\limits^{m}_{\substack{i=0 \\ i \neq l}}(x^l -x^i)},\quad l = 0 \ldots m \, .
\label{eq:analytic}
\end{equation}
The normalization factor $q(m)$ in~(\ref{eq:jumpapprox}) is
\begin{equation}
q_m(x) = \sum_{x^l \in \mathcal{S}^{+}(x)} c_l(x),
\end{equation}
where $\mathcal{S}^{+}(x)$ is the set  \{$x^l : \, x^l \in \mathcal{S}(x), \,  x^l>x$\}. Finally, the accuracy of this approximation is
\begin{equation}
\label{e:PAerror}
    L_mf(x) = \left\{
\begin{array}{l l}
    [f](\xi) + \mathcal{O}(h(x)) & \quad \mathrm{if} \ x^{l-1} \leq \xi,x \leq x^l,\\[6pt]
 \mathcal{O}(h^{\min(m,k)}(x)) & \quad \mathrm{if} \  f \in C^k(I_x) \  \mathrm{for}\  k > 0,\\
 \end{array} \right.
\end{equation}
where $\xi$ is a location at which $f$ has a jump discontinuity, $I_x$ is the smallest interval of points \{$x^l$\} that contains the set $\mathcal{S}(x)$, and $h(x)$ is defined as the largest difference between neighboring points in the stencil $\mathcal{S}({x})$,
\begin{equation}
    h(x) = \max{\{\left|x^i - x^{i-1}\right| : x^{i-1},x^i \in \mathcal{S}(x)\}} \, .
\end{equation}
A proof of (\ref{e:PAerror}) is given in \cite{Archibald2005} and is based on the residual of the Taylor series expansion around the point at which the jump function is being evaluated. Note that the expressions above rely on choosing a particular order of annihilation $m$; as proposed in \cite{Archibald2005}, we use the \emph{minmod} scheme to enhance the performance of polynomial annihilation by evaluating the jump function over a range of orders $\mathcal{M} \ni m$. We will apply the one-dimensional polynomial annihilation scheme along each coordinate direction in order to extend it to multiple dimensions; this process will be detailed in Section~\ref{s:PAinit}. 

\subsection{Support vector machines}
In the algorithm to be detailed in Section~\ref{sec:method}, we will label function evaluations according to which side of the separating surface they lie on. A support vector machine (SVM)~\cite{Burges1998, Scholkopf2002, Vapnik1995}, a supervised learning technique, is then used to build a boundary between the different classes of points. The classification boundary thus becomes an approximation of the separating surface. 

The basic idea behind SVMs is to obtain a function or ``classifier'' of the form:
\begin{equation}\label{eq:svm}
f_{\lambda}^{\ast}(x) = \sum_{i=1}^{N} \alpha_i K(x^i,x),
\end{equation}
where $\alpha_i$ are coefficients associated with locations of the data points $x^i$, $\lambda$ is a regularization parameter, and $K$ is a Mercer kernel~\cite{Mercer1909}. Evaluation of the kernel yields the dot product between two points in a higher-dimensional feature space in which a linear classification boundary is sought. This feature space is the reproducing kernel Hilbert space (RKHS) $\mathcal{H}_K$ induced by the kernel. In other words, one can define the mapping $\Phi: \mathcal{X} \rightarrow \mathcal{H}_K$ and represent the kernel as $K(x,y) = \Phi^T(x)\Phi(x)$. To find the SVM classifier, however, only this inner product is needed. Thus $\Phi$ need not be specified explicitly. This is important because dimensionality of the feature space can be quite large---for example, infinity in the case of a Gaussian kernel. 

The SVM classifer is a solution to a regularized least squares problem with hinge loss given by
\begin{equation}
f_{\lambda}^{\ast}(x) = \arg \min_{f \in \mathcal{H}_K} \left\{ \frac{1}{n}\sum_{i=0}^n \max\left(0, 1-y^if(x^i)\right) + \lambda ||f||^2_{\mathcal{H}_K} \right\},
\end{equation}
where $n$ is the number of training points, $x^i$ are the training points, $y^i$ are the labels of training point $i$, $f(x^i)$ is the classifier function evaluated at training point $i$, and $\lambda$ is a regularization parameter. From this optimization problem we see that the classifier is determined by its sign: $f_{\lambda}^{\ast}(x) > 0$ if $x \in R_1$ and $f_{\lambda}^{\ast}(x) < 0$ if $x \in R_2$, where $R_1$ and $R_2$ are the regions, or classes, bounded by the separating surface. While the sign of the classifier indicates the region/class to which any point belongs, its magnitude reflects the distance a point lies from the boundary in the feature space. Points for which $|f_{\lambda}^{\ast}| < 1$ are said to lie within the margin of the classifer, while larger magnitudes of $f_{\lambda}^{\ast}$ correspond to points increasingly further from the classifier boundary.

Implementation of the SVM involves selecting a kernel. In this work we use a Gaussian kernel $K(x,y) = \exp\left\{-\|x-y\|^2/2\sigma^2 \right\}$. The computational cost of finding the classifier using the SMO algorithm in~\cite{Platt1998} is problem dependent, but can range from $\mathcal{O}(N)$ to $\mathcal{O}(N^2)$ \cite{Platt1998}. Additional costs may be incurred depending on the choice of cross validation techniques to select the parameters involved in the kernel (e.g., $\sigma$) and the penalty on misclassified training samples, $\lambda$. Choosing a small $\sigma$ or a small $\lambda$ can lead to large generalization errors because of overfitting, but choosing large values can cause a loss of complexity of the representation (underfitting). For the algorithm described in this work, LIBSVM~\cite{Chang2011} is used to implement SVMs.

\subsection{Uncertainty sampling and active learning}
Active learning \cite{Cohn1994, Schohn2000, Settles2010} and specifically uncertainty sampling (US) \cite{Catlett1994} are unsupervised learning techniques commonly used in the machine learning community when labeling data points according to their class is an expensive process. In this context one would like to select, from a large unlabeled set, a small subset of points most useful for constructing or refining a classifier; only the selected points are then labeled.
In the discontinuity detection problem, we are free to evaluate the model anywhere in the domain; however, each evaluation is expensive and requires careful selection. Uncertainty sampling involves only evaluating the model in locations where the classifier is relatively uncertain about the class to which a data point belongs. In these situations US is used to add data points adaptively to a data set, retraining the classifier after each addition.

In the context of SVMs, one may define the uncertainty as the closeness of the evaluating point to the boundary. As described above, this closeness is measured by the magnitude of the classifier function~(\ref{eq:svm}). An application of SVMs in this context can be found in the reliability design and optimization literature~\cite{Basudhar2008}, where active learning was used to help refine the boundary of a failure region.


\section{Discontinuity detection algorithm}\label{sec:method}
The algorithm presented in this section takes advantage of any regularity exhibited by the separating surface, avoids the creation of structured grids and nested rectangular sub-domains, and incorporates guided random sampling to improve scaling with dimension. These features of the algorithm result from an integration of the tools discussed in Section~\ref{sec:tools}. Polynomial annihilation is used to obtain general information about the the size and location of the discontinuity, and the regularity of the separating surface is exploited by the SVM classifier. Approximating the separating surface using SVMs allows for a more efficient description of the discontinuity than the nearest neighbor approach used in edge tracking and adaptive refinement schemes. Additionally, SVMs are robust and tend to not overfit the data due to the regularization described above.

\begin{figure}
  \begin{center}
    \input{figs/wflow_wo_approx}
    \caption{Flow chart of the discontinuity detection algorithm.}
    \label{fig:flow}
\end{center}
\end{figure}
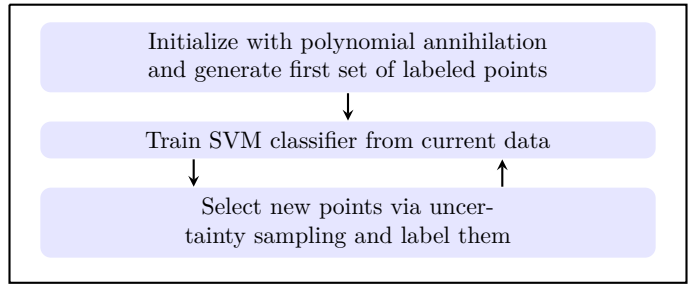

The methodology employed to detect and parameterize the separating surface can be described in three steps, depicted in Figure~\ref{fig:flow}. 
 The first step is an initialization that involves identifying essential characteristics of the discontinuity, such as the jump size and the approximate location of the separating surface, at several points in the parameter domain. This step also seeds the labeling mechanism by which new model evaluations may be classified according to their value. The second and third steps are then alternated repeatedly. The second step involves constructing an SVM classifier to describe the separating surface. The third step involves refining the SVM classifier by selecting new model evaluation points via uncertainty sampling and labeling these points. 

In the remainder of this section, we will use $x \in \mathbb{R}^d$ to denote the $d$-dimensional model parameters. The model is now $f: \mathbb{R}^d \rightarrow \mathbb{R}$. Subscripted variables denote position along the coordinate axes, i.e., $x_j \in \mathbb{R}$ is the $j$th coordinate of $x$, while superscripts are used to index sample points.

\subsection{Initialization with polynomial annihilation}
\label{s:PAinit}
The purpose of an initialization phase of the discontinuity detection algorithm is ultimately to provide a mechanism for labeling future model evaluations according to their values. This labeling is necessary to provide a labeled set of points with which to build the SVM classifiers. Note that polynomial annihilation is used to label points according to their \textit{function value}, whereas the SVM is used to label points based upon their \textit{location} in parameter space. 

The initialization procedure is essentially a divide-and-conquer approach guided by repeated applications of one-dimensional polynomial annihilation. It is similar to the procedure found in~\cite{Jakeman2011}. The procedure begins with an initial set of function evaluations and ends with a set of \textit{jump function} values at various points in the parameter space. These points are surrounded by additional points at which the model (but not the jump function) was evaluated. One major difference between our implementation and that of \cite{Jakeman2011} involves the selection of points used in each PA calculation. In particular, we define an off-axis tolerance \texttt{tol} which is used to define the axial point set $\mathcal{S}(x)$ described in Section~\ref{sec:pa}.  Intuitively, the off-axis tolerance reflects an accepted minimum resolution level of the discontinuity, as described below. 

\begin{figure}[H]
\begin{center}
\input{figs/pic}
\caption{Selection of the set $\mathcal{S}_j(x)$ for performing polynomial annihilation along the horizontal axis $x_j$.}
\label{fig:stencil}
\end{center}
\end{figure}
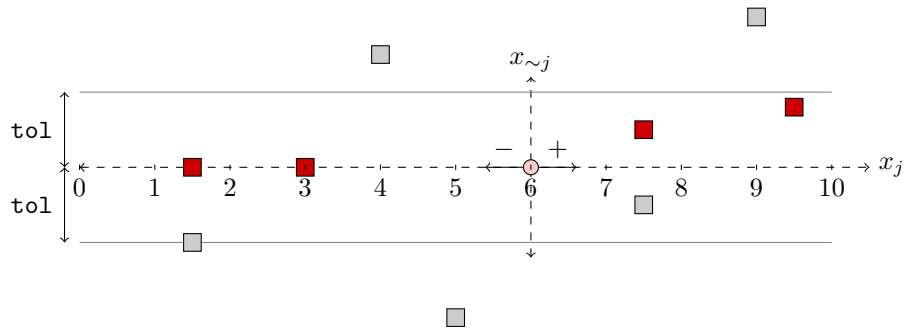

Figure~\ref{fig:stencil} illustrates the application of polynomial annihilation along the horizontal dashed line (the $x_j$ axis), and in particular, our method for choosing the point set $\mathcal{S}_j(x)$ used to perform PA in the $j$th coordinate direction around a point $x$. The vertical dashed line denotes all other coordinate directions, $x_{\sim j} \in \mathbb{R}^{d-1}$. The point of interest or POI $x^{p}$, denoted by the pink circle, is the point at which the jump function will be evaluated. The arrows labeled $+$ and $-$ refer to the relative directions, along the $x_j$ axis, of the surrounding points. For the purposes of polynomial annihilation, at least one point on either side of the POI is necessary. The boxes denote points at which we have performed function evaluations. Two light grey lines bound the region within \texttt{tol} of the axis, wherein all points are considered to be ``semi-axial'' and thus suitable for detecting a discontinuity along $x_j$. In other words, any of the boxes within the grey lines may be considered for the set $\mathcal{S}_j(x^{p})$; those actually selected for this set are drawn in red.

Two special cases are illustrated in Figure~\ref{fig:stencil}. The first special case involves the points located at $x_j=1.5$. These points are equidistant from the POI along the $x_j$ axis, and in this situation the point with the smaller Euclidean distance (in all $d$ directions) from the POI is chosen. The second special case involves the points at $x_j=7.5$. These points are equidistant from the POI both in the $x_j$ direction and in total distance. In this situation either of the points may be chosen; the top point is chosen here for illustration, but the tie is broken randomly in practice. Once the points available for the stencil are determined, the number actually used for PA is determined by the desired annihilation order $m$. Following this selection, we approximate the jump function at the POI using (\ref{eq:jumpapprox}). 

The inclusion of semi-axial points in $\mathcal{S}_j(x^{p})$ may affect the accuracy of the jump function approximation. We can analyze this effect by considering the error induced by incorrectly evaluating the function values $f(x^l)$ in (\ref{eq:jumpapprox}). Suppose that we are trying to approximate the jump function at a point $x$ in direction $j$ using semi-axial neighbors $\hat{x}^l \in \mathcal{S}_j(x)$ and $\texttt{tol}>0$. We perform the polynomial annihilation procedure as if we have truly on-axis points given by $x^l := (x_1, \ldots, x_{j-1}, \hat{x}^l_j, x_{j+1}, \ldots x_{d})$, one for each element of $\mathcal{S}_j(x)$. The jump function approximation is computed as
\begin{equation}
  \label{eq:approxjumpapprox}
  \widetilde{L}_m f(x) = \frac{1}{q_m(x)} \sum_{\hat{x}^l \in \mathcal{S}_j(x)} c_l(x) f(\hat{x}^l),
\end{equation}
where we recall that $c_l$ and $q_m$ depend only on the $j$th coordinate of the points in $\mathcal{S}_j(x)$, and hence their values are equivalent for $x^l$ and $\hat{x}^l$. The difference between the approximation above \eqref{eq:approxjumpapprox} and that in \eqref{eq:jumpapprox} is then due only to evaluating $f$ at $\hat{x}^l$ rather than at $x^l$:
\begin{equation}
  \widetilde{L}_mf(x) -L_mf(x) = \frac{1}{q_m(x)} \sum_{\hat{x}^l \in \mathcal{S}_j(x)} c_l(x) \left[ f(\hat{x}^l) - f(x^l)\right] .
\end{equation}
If a discontinuity exists in between $\hat{x}^l$ and $x^l$, then errors introduced into the approximation will be on the order of the size of the jump. However, if for every $\hat{x}^l$, $f$ is continuous on the closed interval between $\hat{x}^l$ and $x^l$ and differentiable on the corresponding open interval, then we can use the mean value theorem to bound the magnitude of the difference $|f(\hat{x}^l) - f(x^l)|$. Under these conditions, for each $l$ there exists a $\nu^l = (1-\eta^l)x^l + \eta^l \hat{x}^l$ with $\eta^l \in (0,1)$ such that
\begin{equation}
  f(\hat{x}^l) - f(x^l) = \sum_{i=1, i \neq j}^d \partial_{x_i} f(\nu^l)
\left( \hat{x}^l_i - x^l_i \right). \label{eq:meanvalue}
\end{equation}
Then
\begin{equation}
  |f(\hat{x}^l) - f(x^l)| \leq \sum_{i=1, i \neq j}^d \left| \partial_{x_i} f(\nu^l) \right| \left| \hat{x}^l_i - x^l_i \right| \leq \texttt{tol} \sum_{i=1, i \neq j}^d \left| \partial_{x_i} f(\nu^l)\right |. 
\label{eq:magdiff}
\end{equation}
Now let $G = \max \limits_{l} \max\limits_{i \neq j} \sup\limits_{s \in \mathcal{B}_j(x^l)} \left| \partial_{x_i} f(s) \right |$, where $\mathcal{B}_j(x^l)$ is a ball of radius $\texttt{tol}$ surrounding $x^l$ in the $x_{\sim j}$ directions, i.e., $\mathcal{B}_j(x^{l}) := \left\{ (s_1, \ldots, s_{j-1}, x^l_{j}, s_{j+1}, \ldots, s_d): |s_i - x^l_i | < \texttt{tol}, \, i=1\ldots d, \, i \neq j\right\}$. Then we can bound the difference \eqref{eq:magdiff} above  by $G (d-1) \, \texttt{tol}$.  The magnitude of the difference between \eqref{eq:jumpapprox} and \eqref{eq:approxjumpapprox} can then be estimated as
\begin{equation}
  \left| \widetilde{L}_m f(x) -L_m f(x) \right|  \leq  
  \frac{1}{| q_m(x) |} \sum_{\hat{x}^l \in \mathcal{S}_j(x)} | c_l(x) | \left | f(\hat{x}^l) - f(x^l)\right| = 
 \mathcal{O}(G d\, \texttt{tol}) ,
\end{equation}
where the second step uses the fact that both $c_l(x)$ and $q_m(x)$ are of the same magnitude, $\mathcal{O}(h(x)^{-m})$ \cite{Archibald2005}. An application of the triangle inequality then yields an update to the error estimate (\ref{e:PAerror}) for approximation of the jump function:
\begin{equation}
  \label{e:PAerrorApprox}
  \widetilde{L}_mf(x) = \left\{
    \begin{array}{l l}
        [f](\xi) + \mathcal{O}(h(x)) + \mathcal{O}\left(G d \, \texttt{tol}\right) & \quad \mathrm{if} \ \hat{x}_j^{l-1} \leq \xi, x_j \leq \hat{x}_j^l,\\[6pt]
     \mathcal{O}(h^{\min(m,k)}(x)) + \mathcal{O}\left(G d \, \texttt{tol}\right) & \quad \mathrm{if} \  f \in C^k(I_x) \  \mathrm{for}\  k > 0.
     \end{array} \right. 
    \end{equation}
In this multidimensional case, $\xi \in \mathbb{R}^d$ but differs from $x$ only in dimenson $j$, i.e., $\xi_i = x_i$ for $i \neq j$, and $\xi_j$ is a location of a jump discontinuity along the $x_j$ axis. $I_x$ and $h(x)$ are defined just as in Section~\ref{sec:pa}, using only the $j$th coordinates of the points in the set $\mathcal{S}_j(x)$. This simple estimate suggests that as long as $G  d \, \texttt{tol}$ is significantly smaller than the jump size $[f](\xi)$, only small errors will be induced in the jump function approximation by using off-axis points. Intuitively this means that the off-axis tolerance should be kept small enough to balance the variation of the function in the off-axis directions.

Now that we have described the selection of points used for each one-dimensional application of PA, we the describe the multi-dimensional initialization procedure. This algorithm is based on a repeated divide-and-conquer refinement of some initial set of points. Each refinement further localizes the discontinuity. The core of the divide-and-conquer approach for PA requires the evaluation of the jump function at various test points based on a set of previously evaluated data points. 
The algorithm is recursive in the sense that at any given step, we wish to refine the location of the discontinuity in direction $j$ and at a given location $x$. We do this by first finding two additional points at which to evaluate the jump function; these points, $y^1$ and $y^2$, are chosen to be the midpoints between $x$ and its nearest semi-axial neighbors in the $\pm j$ directions. Next, we evaluate the jump function at each of these locations, $J^1=[f](y^1)$ and $J^2=[f](y^2)$. If the value of $J$ indicates that a jump exists at either of these points (i.e., up to the accuracy given in (\ref{e:PAerror})), then we either evaluate the full model $f$ at the point and perform the same procedure recursively in every other coordinate direction, or we add the point to the set of edge points $\mathcal{E}$ and stop refining around it. 
Before recursively performing the procedure in a particular direction $k$ for a point $y$, we evaluate $f$ on the $k$-semi-axial \textit{boundary parents} corresponding to $y$, if these evaluations do not already exist. These boundary parents are locations on the boundary of the parameter space in the +$k$ and -$k$ directions.\footnote{If the parameter space is unbounded, then the boundary parents can be any $k$-semi-axial points that are far enough from $y$ in the $k$ direction to ensure that the stencil for jump function evaluation at $y$ is not too narrow.} Once the function is evaluated at these parent locations, we are assured to have a sufficient number of semi-axial function evaluations to perform PA. The set of edge points $\mathcal{E}$ is the set of points at which we have found nonzero approximations of the jump function and that are located within an \textit{edge tolerance} $\delta$ of two other points at which we have evaluated the function. The entire algorithm exits when either no more refinement is possible or the cardinality of the set of edge points reaches a user defined value $N_{E}$.

Algorithm~\ref{alg:init}, \texttt{RefinementInitialization}, initializes the refinement of the discontinuity by calling Algorithm~\ref{alg:2}, \texttt{Refine1D}, for each initial point in a set $\mathcal{M}_0$. In practice, we often start either with a single point at the origin, or randomly sampled points through out the regime. Initialization with randomly sampled points can provide a more robust method for finding the separating surface since they force an exploration of a wider area of the parameter domain.
\texttt{Refine1D} recursively refines the location of the discontinuity as described above. Both are detailed below, and constitute the PA phase of the overall discontinuity detection algorithm. For reference, the function $\mathrm{NN}_{\pm k}(\mathcal{S},x)$ finds the nearest neighbor to the point $x$ in the $\pm k$ coordinate direction, among the points in the set $\mathcal{S}$.

\begin{algorithm}
\caption{\texttt{RefinementInitialization}}
\label{alg:init}
\begin{algorithmic}[1]
\STATE \textbf{Input:} initial point set $\mathcal{M}_0$; maximum number of edge points $N_E$; edge tolerance $\delta$; off-axis tolerance $\mathtt{tol}$
\STATE \textbf{Initialize:} $\mathcal{M} = \mathcal{M}_0$, $\mathcal{E} = \emptyset$, $ \mathcal{F} = \{f(x^1), f(x^2), \ldots, f(x^n) :  x^i \in \mathcal{M} \}$
\FORALL {$x^i \in \mathcal{M}$}
	\FOR {$j$ from $1$ to $d$}
        \STATE If needed, add boundary parents of $x^i$ in direction $j$ and their function values to $\mathcal{M}$ and $\mathcal{F}$ respectively
        \STATE ($\mathcal{M}, \mathcal{E}, \mathcal{F}$) = \texttt{Refine1D}($\mathcal{M}$, $\mathcal{E}$, $\mathcal{F}$, $x^i$, $j$, $N_E$, $\delta$, $\mathtt{tol}$)
            \IF {$|\mathcal{E}| \geq N_{E}$}
                \STATE Return ($\mathcal{M}, \mathcal{E}, \mathcal{F}$)
            \ENDIF
        \ENDFOR
    \ENDFOR
\STATE Return ($\mathcal{M}, \mathcal{E}, \mathcal{F}$)
\end{algorithmic}
\end{algorithm}

\begin{algorithm}
\caption{\texttt{Refine1D}}
\label{alg:2}
\begin{algorithmic}[1]
\STATE \textbf{Input:} point set $\mathcal{M}$; edge point set $\mathcal{E}$; model evaluations $\mathcal{F}$; location for refinement $x$; coordinate direction of refinement $j$; maximum number of edge points $N_E$; edge tolerance $\delta$; off-axis tolerance $\mathtt{tol}$
    \STATE Determine $\mathcal{S} := \mathcal{S}_j(x)$ using tolerance $\mathtt{tol}.$
    \STATE Define $y^1 = \left(x + \mathrm{NN}_{+j}\left(S, x\right) \right)/2$
    \STATE Define $y^{2} = \left(x + \mathrm{NN}_{-j}\left(S, x\right) \right)/2$
	\STATE $J^1 = [f](y^1)$
    \STATE $J^2 = [f](y^2)$
	\FOR {each $k \in {1,2}$}
        \IF {$J^k$ indicates jump exists}
		    \IF{ $\|y^k - x\| \leq \delta$}
			    \STATE Add $y^k$ to $\mathcal{E}$.
                \IF {$|\mathcal{E}| \geq N_{E}$}
                    \STATE Return ($\mathcal{M}, \mathcal{E}, \mathcal{F}$)
                \ENDIF
		    \ELSE
			    \STATE Add $y^k$ to $\mathcal{M}$.
			    \STATE Add $f(y^k)$ to $\mathcal{F}$.
                \FOR {$l$ from $1$ to $d$}
                    \STATE If needed, add boundary parents of $y^k$ in direction $l$ and their function values to $\mathcal{M}$ and $\mathcal{F}$ respectively
                    \STATE ($\mathcal{M}, \mathcal{E}, \mathcal{F}$) = Refine1D($\mathcal{M}$, $\mathcal{E}$, $\mathcal{F}$, $y^k$, $l$, $N_E$, $\delta$, $\mathtt{tol}$)
                \ENDFOR
		    \ENDIF
	    \ENDIF
    \ENDFOR
    \STATE Return ($\mathcal{M}, \mathcal{E}, \mathcal{F}$)
\end{algorithmic}
\end{algorithm}

\subsection{Labeling in the initialization phase}\label{sec:labelPA}
Having estimated the jump size and location of the discontinuity at a few points in the parameter space using polynomial annihilation, we would like to use these estimates to \textit{label} the points in $\mathcal{M}$ according to the function evaluations already performed by the initialization procedure (and stored in $\mathcal{F}$). Determining the class in which a point resides depends on the jump values at the edge points. We only label the points in $\mathcal{M}$ that lie within the edge tolerance $\delta$ of points in $\mathcal{E}$. Recall that each edge point (i.e., each element of $\mathcal{E}$) lies within $\delta$ of at least two points in $\mathcal{M}$. 

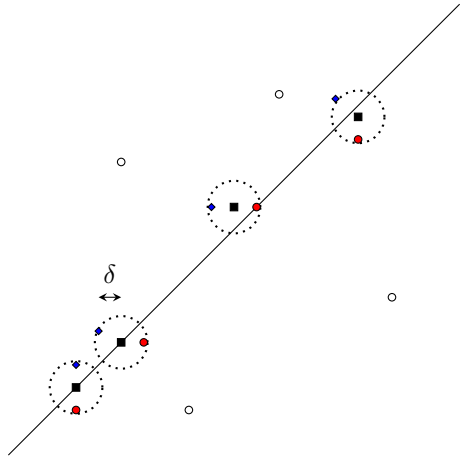
\begin{figure}
\begin{centering}
\input{figs/labelex}
\caption{Polynomial annihilation-based labeling procedure, used during the initialization phase of the algorithm. Circles and diamonds are locations where the function has been evaluated. Squares are edge points. Blue diamonds are function evaluations that are labeled as class 1 and red circles are function evaluations that are labeled as class 2. The dotted circles are of radius $\delta$.}
\label{fig:label}
\end{centering}
\end{figure}

For each point $y \in \mathcal{E}$, we find the elements of $\mathcal{M}$ within $\delta$ of $y$; call these $\mathcal{M}_y = \{ x :\, x \in \mathcal{M}, |x-y| < \delta \}$.  Of this subset, the point $x^\ast$ with the largest function value is found and labeled \textit{class 1}. Then the function values at all the other points $x \in \mathcal{M}_y$ are compared to $f(x^\ast)$ using the jump value $[f](y)$ as a reference. If the difference between $f(x)$ and $f(x^\ast)$ is less than the jump value, then the point $x$ is labeled \textit{class 1}; otherwise, it is labeled \textit{class 2}. We note that \textit{class 1} therefore always contains the locally largest values by definition. (If it is known a priori that the locally largest values in different parts of the domain should be in different classes, a different labeling procedure that incorporates this knowledge must be used.) The present procedure can successfully label points along a discontinuity whose jump size varies along the domain, without any manual intervention. Figure~\ref{fig:label} illustrates the procedure.

\begin{figure}
\begin{center}
\subfigure[]{\includegraphics[scale=0.3]{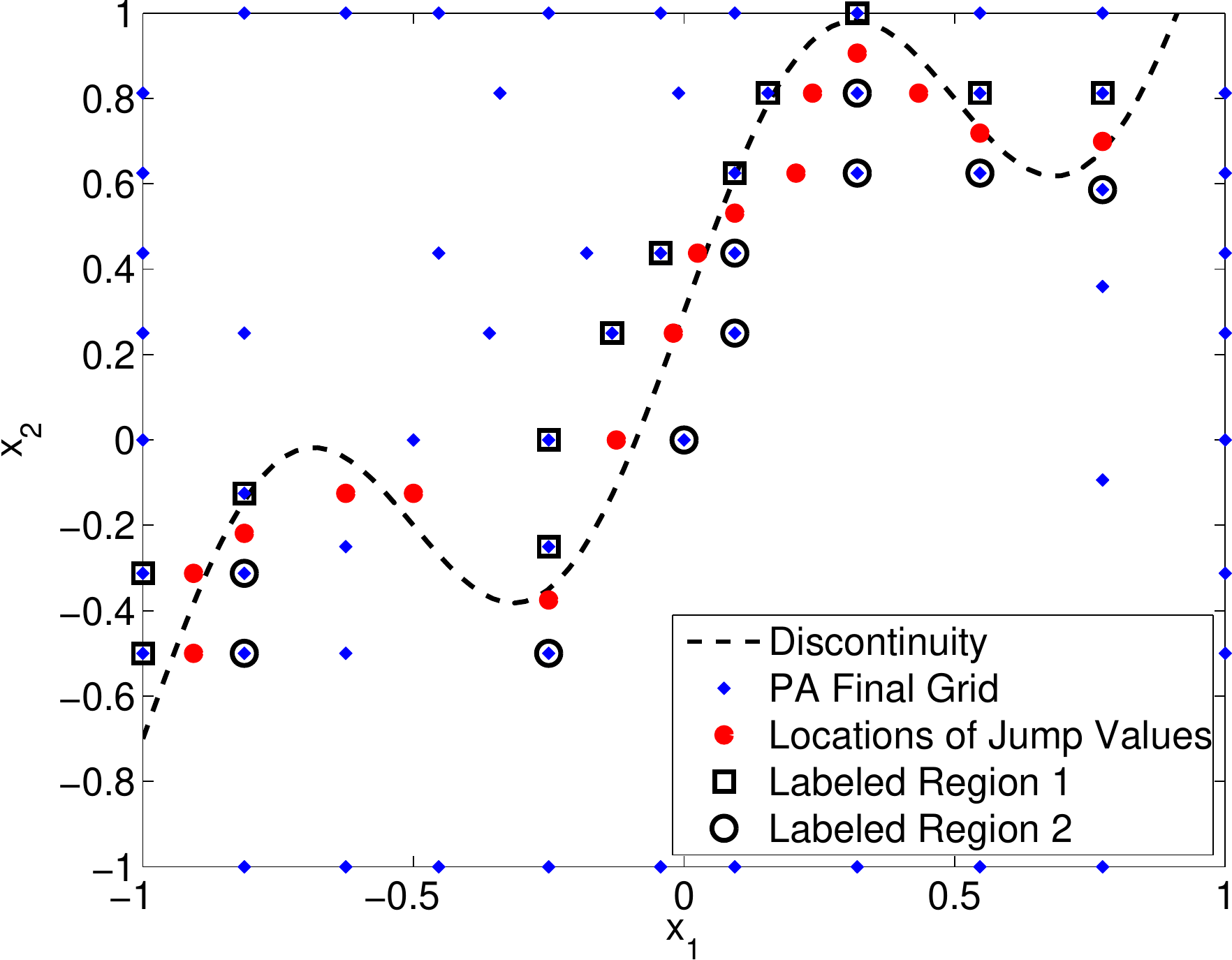} } 
\subfigure[]{\includegraphics[scale=0.3]{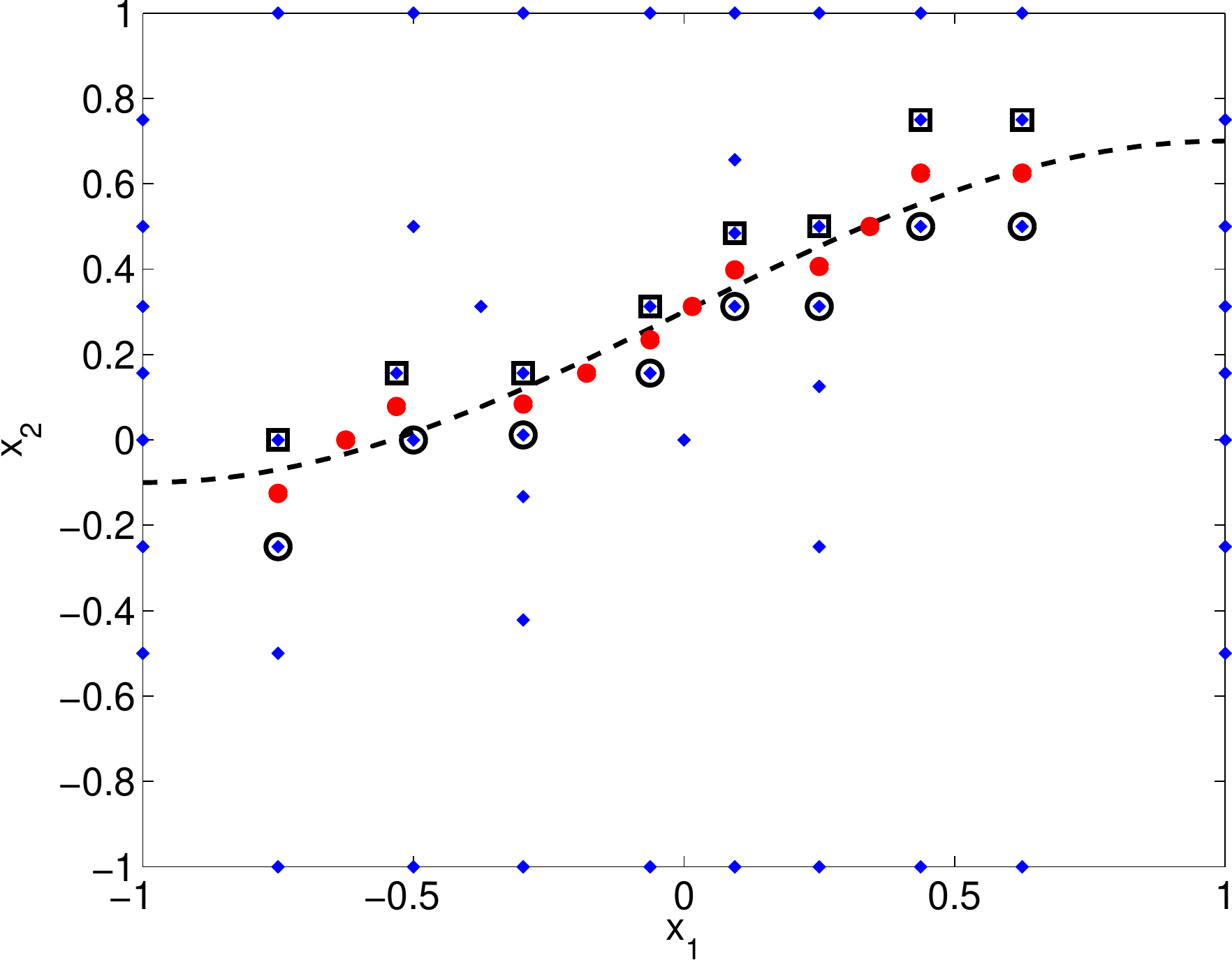} }
\subfigure[]{ \includegraphics[scale=0.3]{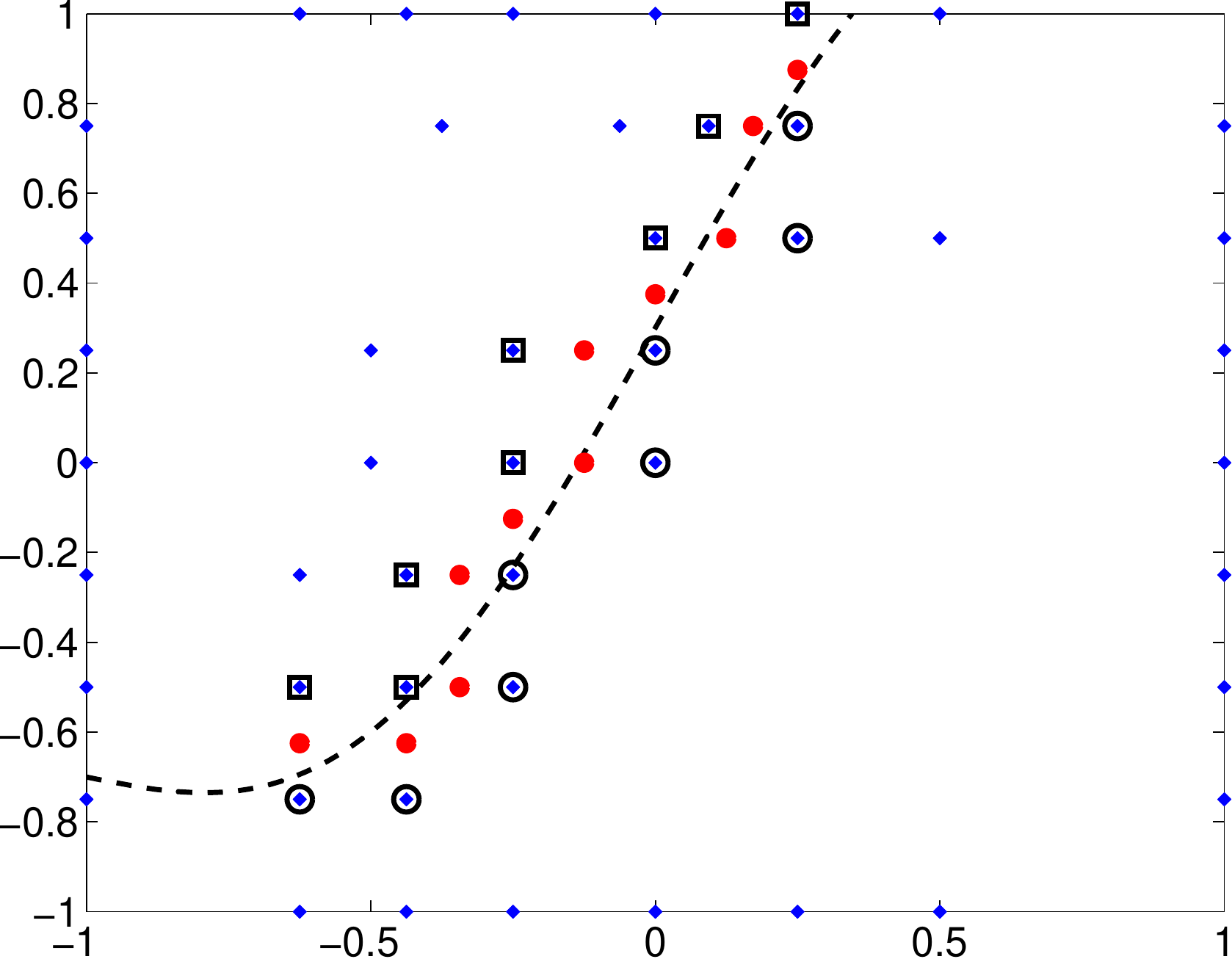}}
\caption{Initialization phase of the discontinuity detection approach (Algorithm~\ref{alg:init}), applied to several discontinuities. All model evaluation points are marked with diamonds; labeled points have an additional circle or square. Edge points are indicated with red circles.}
\label{fig:PAapplied}
\end{center}
\end{figure}

The complete initialization phase of the algorithm is now demonstrated on three test discontinuities, shown in Figure~\ref{fig:PAapplied}. In these numerical experiments, $\delta$ and $\mathtt{tol}$ are both set at 0.125 and $\mathcal{M}_0$ consists of a single point at the origin. In these plots we see that the discontinuity is located and surrounded by a very coarse grid of function evaluations. The red circles points indicate locations at which we obtain jump values approximating the size of the discontinuity. These are the points in set $\mathcal{E}$, used to label the surrounding function evaluations as described above.

\subsection{Refinement with uncertainty sampling}\label{sec:uslabel}

We now describe the use of SVM classification and active learning to refine our description of the discontinuity, following the initialization phase of the algorithm, i.e., Algorithm~\ref{alg:init}. Compared to simply continuing Algorithm~\ref{alg:init} with smaller tolerances to generate more edge points, this phase of the algorithm focuses on choosing model evaluation points that are most informative for the SVM classifier. Later we will demonstrate, via numerical examples in Section~\ref{sec:scaling}, that switching to the active learning phase after relatively few iterations of Algorithm~\ref{alg:init} results in significant efficiency gains and improved scaling of computational effort with parameter dimension. The steps described in this section comprise the second and third boxes in the flow chart of Figure~\ref{fig:flow}: SVM classification, using all of the currently labeled points (\textit{class 1}, \textit{class 2}), alternates with the selection of new points via uncertainty sampling and the labeling of these new points.  

New points are chosen based on their proximity to the zero level set of the current classifier. We obtain a point near the classifier boundary by first \textit{drawing a sample} from an underlying measure on $x$ (e.g., a probability measure on the input parameters of the model), and then using this sample as an initial guess for the following optimization problem, which minimizes the square of the classifier function in~(\ref{eq:svm}) and thus drives the initial guess towards the boundary:
\begin{equation}
\label{eq:UScostfunc}
\min_x \left(\sum_{i=1}^{N}\alpha_i K(x^i, x)\right)^2  .
\end{equation}
Here $\{x^1,\ldots, x^N\}$ are the support vectors of the current classifier. A variety of optimization algorithms can be used for this purpose and result in similar performance. Note that this optimization problem is not convex and may have multiple local minima. But the initial randomization helps mitigate clustering of points in local minima, and in practice the possibility of clustering does not impede refinement of the discontinuity; as the SVM is updated, these minima are themselves altered. Moreover, we compel the generation of ``low discrepancy'' points along the discontinuity by constraining new function evaluations to occur farther than a minimum distance $\epsilon$ from existing points. In other words, a candidate point $x^\ast$ found by minimizing~\ref{eq:UScostfunc} is not evaluated or used for classification if it lies less than $\epsilon$ away from the nearest evaluated node. 
As uncertainty sampling progresses, the entire discontinuity will be explored with resolution $\epsilon$. Eventually the algorithm results in randomly distributed training points that approximate a Monte Carlo edge tracking scheme. The uncertainty sampling scheme is precisely detailed in Algorithm \ref{alg:findPointsOnClassifier}.

\begin{algorithm}
\caption{\texttt{FindPointsOnClassifierBoundary}}
\label{alg:findPointsOnClassifier}
\begin{algorithmic}[1]
\STATE \textbf{Input:} set $\mathcal{L} = \left\{ (x, \ell) \right\}$ of (points, labels); number of points to add $N_{add}$; variation radius $\delta_t$; resolution level $\epsilon < \delta_t$; maximum number of iterations \texttt{itermax}
\STATE $N_{added} = 0$
\STATE  $\mathcal{X} = \emptyset$
\STATE $\mathtt{iter} = 1$
\WHILE {$N_{added} < N_{add}$  and $\mathtt{iter} < \mathtt{itermax}$}
	\STATE $x  \leftarrow $ \texttt{samplePointFromDomain()} \COMMENT{Equation~(\ref{eq:UScostfunc})}
	\IF  {$\arg \min_{(x^\ast,\ell^\ast) \in \mathcal{L}}|| x - x^\ast|| > \epsilon$}
		\IF { $\exists  (x^1,\ell^1) \in  \mathcal{L}$ \, s.t.\ $||x^1-x|| < \delta_t$ and $\ell^1=+1$ }
			\IF { $\exists  (x^2,\ell^2) \in  \mathcal{L} $ \, s.t.\ $||x^2-x|| < \delta_t$ and $\ell^2=-1$ }
				\STATE $N_{added} = N_{added}+1$
				\STATE $\mathcal{X} \leftarrow \mathcal{X} \cup \left\{ x \right\}$
			\ENDIF
		\ENDIF
	\ENDIF
	\STATE {$\mathtt{iter} = \mathtt{iter}+1$}
\STATE Return($\mathcal{X}$)	
\ENDWHILE
\end{algorithmic}
\end{algorithm}

When a new data point is generated through uncertainty sampling, it must be assigned to a class. Our labeling scheme for the points generated during polynomial annihilation relied on estimates of the jump function, and thus only applied to points within $\delta$ of an edge point. Now during uncertainty sampling, we must label points that potentially lie much further away from edge points, where no local value of the jump function is available. We thus employ a different labeling scheme that compares new points generated during uncertainty sampling to the nearest previously labeled points in each class. 

\begin{figure}
\begin{centering}
\input{figs/labelUS2}
\caption{Labeling procedure for points generated during uncertainty sampling. A new test point (green square) is labeled by comparing its function value to those of its nearest neighbors in Class 1 and Class 2, but only if the test point is sufficiently close to both neighbors. The blue diamonds and red circles denote previously labeled points from Class 1 and 2, respectively. Circles $C_1$ and $C_2$ indicate the regions for each class in which the local variability of the function is small enough for labeling. The radii of these circles are $\delta_t^1$ and $\delta_t^2$ and are chosen based on the discussion in Section~\ref{sec:uslabel}. New points within the intersection of these circles can be accurately labeled.} 
\label{fig:labelUS}
\end{centering}
\end{figure}
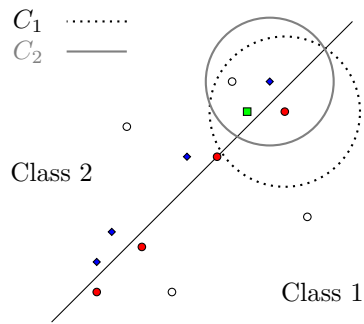

In particular we define a new tolerance $\delta_t$ which reflects the radius of a region in each class within which the \textit{local} variability of the function along the separating surface is smaller than the local jump size itself. In principle we can specify a separate $\delta_t$ for each class, but for simplicity we consider the same value for both. A new point is now labeled only if its nearest neighbors in each class are located within a distance $\delta_t$. The function value at the new point is compared to the function value of its nearest neighbor in \textit{class 1} and its nearest neighbor in \textit{class 2}. The new point is given the same label as the nearest neighbor with the closest function value. For this scheme to avoid making any errors, $\delta_t$ must be chosen properly. We explain this requirement and precisely define the notion of ``local'' as follows.
Suppose that we are attempting to label a new point $x^u$ which has function value $f(x^u)$, and that its nearest neighbors in \textit{class 1} and \textit{class 2} are $x^{(1)}$ and $x^{(2)}$, respectively. Suppose also that both nearest neighbors are within $\delta_t$ of $x^u$. Based on the class definitions in Section~\ref{sec:labelPA}, we can assume that $f(x^{(1)}) > f(x^{(2)})$. 
We now determine the consequences of our labeling mechanism if $x^u$ lies in \textit{class 1}. There are three possible orderings of $f(x^u)$ relative to $f(x^{(1)})$ and $f(x^{(2)})$. If $f(x^u) > f(x^{(1)}) >f(x^{(2)})$, then our scheme will generate the correct label, because the function value of the nearest $\textit{class 1}$ point is closer to that of the new point. If $f(x^u) < f(x^{(2)}) < f(x^{(1)})$, then we will generate an incorrect label; in this situation, the variation of the function \textit{within class 1, near the discontinuity} exceeds the jump size $|f(x^{(1)})-f(x^{(2)})|$. The points $x^{(1)}$ and $x^{(2)}$ are too far from $x^u$ to be useful for labeling, and thus $\delta_t$ has been chosen too large. The final possible ordering is $f(x^{(2)}) < f(x^u) < f(x^{(1)})$. 
In this case, we can still label the point correctly if $f(x^{(1)}) - f(x^u) < \frac{1}{2}|f(x^{(1)}) - f(x^{(2)})|$. Alternatively, if $x^u$ belonged to \textit{class 2}, we would need $f(x^u) - f(x^{(2)}) < \frac{1}{2}|f(x^{(1)}) - f(x^{(2)})|$. To ensure that the appropriate inequalities hold, the radius $\delta_t$ must be specified so that the variation of the function around the new point in each class is smaller than $\frac{1}{2}|f(x^{(1)}) - f(x^{(2)})|$. This radius reflects a region \textit{within} a given class in which the function varies a small amount relative to the jump size. If the radius is any larger, this labeling procedure may not be accurate for this final ordering. 
If the jump size is large relative to the local variation of the function throughout the parameter domain, then $\delta_t$ can be quite large, even infinity. If the function values near the separating surface within a particular class vary widely, however, then a smaller $\delta_t$ is needed to ensure accurate labeling. In order to conservatively choose $\delta_t$, one may set $\delta_t = \delta$, i.e., the same as the edge tolerance. In this situation US begins labeling new points which are near existing labeled PA points. These labels will be correct because these points are in a region where we have an accurate approximation of the jump size. Alternatively, one may choose to perform PA with a larger total number of edge points $N_E$ and/or a smaller edge tolerance $\delta$. These changes would yield a more extensive exploration of the separating surface in the PA initialization phase of the algorithm, obtaining jump value estimates at more areas along the separating surface. The main consequence of setting $\delta_t$ too small or performing a large amount of PA is a loss in efficiency; only samples very close to existing samples will be labeled, and progress along the discontinuity will be slow. But the conditions favoring large $\delta_t$ are likely to be valid in practice, as many discontinuities in problems of interest involve relatively large jumps. An example of this labeling procedure is shown in Figure~\ref{fig:labelUS}, where the radius $\delta_t$ is different in each class for illustration purposes. In practice we specify $\delta_t$ to be equal in each class.

Once the new point is labeled, a new classifier is trained and the procedure is repeated: uncertainty sampling, labeling, and SVM training. If a sufficient number of function evaluations are added, this procedure will revert to a Monte Carlo edge tracking scheme with resolution level $\epsilon$, with the SVM interpolating the separating surface among the support vectors. Uncertainty sampling can also be used to add several new points at a time, by minimizing (\ref{eq:UScostfunc}) from several starting points at each iteration. This does not typically reduce the number of function evaluations needed to refine the discontinuity, but can lower overall computational cost by reducing the number of classifiers to be trained. 

\subsection{Stopping criterion}
The stopping criterion for the overall discontinuity detection algorithm is specified by the distance tolerance $\epsilon$ described above (i.e., the distance used to reject too-closely spaced points during uncertainty sampling). For a given value of $\epsilon$, eventually a sufficient number of points are added so that any additional point along the discontinuity lies within $\epsilon$ of a previously labeled point;\footnote{In the case of an unbounded parameter domain endowed with finite probability measure, this will only hold true with high probability.} at this stage the algorithm exits. In practice, the exit criterion is implemented by making many repeated attempts at adding new samples, and exiting after a specified number of failed attempts occur in sequence. In Algorithm~\ref{alg:findPointsOnClassifier} this number of attempts is given by $\texttt{itermax}$.

We also considered using cross validation as a stopping criterion, but found it to be inadequate for this purpose. The reason is that cross validation can only indicate whether points that are \textit{already labeled} are correctly classified by the SVM. These points are not distributed randomly over the entire domain; rather, they are clustered around areas of discontinuity that have already been identified. Cross validation thus has no means of revealing whether the entire discontinuity has been explored. The only way to be sure that the entire discontinuity has been explored is to add points until a desired resolution level is achieved along the entire discontinuity, namely $\epsilon$.

Using this stopping criterion and running the algorithm to completion, the number of points required will grow exponentially with the dimension of the separating surface. Even though the algorithm involves random sampling, the stopping criterion essentially expresses the desire to achieve a space-filling design along the separating surface. In practice, however, we have found that stopping well short of a small $\epsilon$ still leads to good results. This behavior can be attributed to random sampling along the separating surface. Random sampling allows wide regions of the discontinuity to be explored asynchronously, with classification errors resulting from gaps between the samples. Because the exploration is asynchronous and spatially distributed, it is easier to stop the algorithm anytime. These exploration contrasts with edge tracking, where one progressively adds samples by `walking' along the separating surface; here, one cannot stop short because large regions of the separating surface have not yet been explored.

The full discontinuity detection algorithm is summarized in Algorithm~\ref{alg:disc_detect}. It takes as inputs an initial point set $\mathcal{M}_0$, the desired number of edge points $N_{E}$, a PA edge tolerance $\delta$, a PA off-axis tolerance \texttt{tol}, a function variation tolerance $\delta_t$ for US labeling, the number of points to add with every iteration of uncertainty sampling $N_{add}$, an uncertainty sampling resolution level $\epsilon$, the maximum number of US sub-iterations \texttt{itermax}, and finally a maximum run time $T$. The algorithm returns the classifier function $f^\ast_{\lambda}$.

\begin{algorithm}
\caption{\texttt{Discontinuity Detection}}
\label{alg:disc_detect}
\begin{algorithmic}[1]
\STATE \textbf{Input:} initial point set $\mathcal{M}_0 = \{x^1, x^2, \ldots, x^n | x^i \in \mathbb{R}^d \}$; maximum number of edge points $N_E$; PA edge tolerance $\delta$; PA off-axis tolerance \texttt{tol}; US variation radius $\delta_t$; number of US points added each iteration $N_{add}$; US resolution level $\epsilon$; maximum number of US sub-iterations \texttt{itermax};  maximum runtime $T$
\STATE $\mathcal{M}, \mathcal{E}, \mathcal{F}$ = \texttt{RefinementInitialization}($\mathcal{M}_0, N_E$, $\delta$, \texttt{tol})
\STATE $\left\{ (x, \ell) \right\}$ = \texttt{generateLabels}($\mathcal{M}, \mathcal{E}, \mathcal{F}$) \COMMENT{Figure~\ref{fig:label}}
\STATE $f_{\lambda}^\ast(x)$ = \texttt{trainSVMClassifier}($\left\{ (x , \ell ) \right\}$)
\WHILE{\texttt{Runtime} $< T$}
	\STATE $\{x^{new}\} =$ \texttt{findPointsOnClassifierBoundary}($\left\{ (x , \ell ) \right\}, N_{add}$, $\delta_t$, $\epsilon$, \texttt{itermax})
	\IF { $\{x^{new}\} = \emptyset$}
		\STATE Return($f_{\lambda}^\ast$)
	\ENDIF
	\STATE $\{y^{new}\} = f( \{x^{new} \})$
	\STATE $\{\ell^{new}\} =$ \texttt{label}($\{x^{new}\}$, $\{y^{new}\}$) \COMMENT{Figure~\ref{fig:labelUS}}
	\STATE $\left\{ (x , \ell ) \right\} \rightarrow \left\{ (x , \ell ) \right\} \cup \left\{ (x^{new}, \ell^{new}) \right\}$
	\STATE $f_{\lambda}^\ast =$ \texttt{trainSVMClassifier}($\left\{ (x , \ell ) \right\}$)
\ENDWHILE
\STATE Return($f_{\lambda}^\ast$)
\end{algorithmic}
\end{algorithm}

%% file: figs/wflow_wo_approx.tex


\tikzstyle{information text} = [rectangle, rounded corners, fill=blue!10,inner sep=1ex, minimum size=4mm, text width=25em, text centered ]
\tikzstyle{flow}=[style=information text,scale=0.9]
\begin{tikzpicture}[>=stealth,thick,]
	
	\draw (-4.5cm,1.0cm)--(4.5cm,1.0cm)--(4.5cm,-2.7cm)--(-4.5cm,-2.7cm) -- (-4.5cm,1.0cm);
	\node[flow,] at (0,0.3cm) (PA) {Initialize with polynomial annihilation and generate first set of labeled points};
	\node[flow] at (0,-0.8cm)(SVM) {Train SVM classifier from current data};
	\node[flow] at (0,-1.9cm)(US) {Select new points via uncertainty sampling and label them};
	\draw [->] (PA) to (SVM);
    \path (SVM.south) -- (SVM.south west) coordinate[pos=0.5] (SVM1);
    \path (US.north) -- (US.north west) coordinate[pos=0.5] (US1);
	\draw [->] (SVM1) to (US1);

    \path (SVM.south) -- (SVM.south east) coordinate[pos=0.5] (SVM2);
    \path (US.north) -- (US.north east) coordinate[pos=0.5] (US2);
	\draw [->] (US2) to (SVM2);




\end{tikzpicture}


%% file: figs/pic.tex


\tikzstyle{gridpoint}=[rectangle,draw,fill=black!20]
\tikzstyle{axpoint}=[rectangle,draw,fill=black!20!red]
\tikzstyle{intpoint}=[circle,draw,fill=red!20,inner sep=0pt, minimum size=2mm]
\begin{tikzpicture}
	\node at (10.8, 0.0)[] {$x_j$};
	\draw[dashed, <->] (0,0) -- +(10.5,0);
	\draw[dashed ,<->] (6, -1.2) -- +(0,2.4);
	\node at (6, 1.4)[] {$x_{\sim j}$};
	\draw[very thin,gray] (0,1) -- ++(10,0);
	\draw[very thin,gray] (0,-1) -- ++(10,0);
	\node at (6,0) (refpoint) [intpoint] {};
	\node at (9,2)		[gridpoint]{};
	\node at (5,-2)		[gridpoint]{};
	\node at (3,0)		[axpoint]{};
	\node at (9.5,0.8)	[axpoint]{};
	\node at (1.5,0)	[axpoint]{};
	\node at (1.5,-1)	[gridpoint]{};
	\node at (7.5,0.5)	[axpoint]{};
	\node at (7.5,-0.5)	[gridpoint]{};
	\node at (4,1.5)	[gridpoint]{};
	\draw [->] (refpoint.west) --  node [above] {$-$} + (-0.5,0);
	\draw [->] (refpoint.east) --  node [above] {$+$} +(0.5,0);
	\draw [<->] (-0.2,0) -- node[left=1pt] {\texttt{tol}} ++(0,1);
	\draw [<->] (-0.2,0) -- node[left=1pt] {\texttt{tol}} ++(0,-1);
	\foreach \x in { 0, 1, 2, 3, 4, 5, 6, 7, 8, 9, 10}
		\draw[xshift=\x cm] (0pt,1pt)--(0pt,-1pt) node[below] {$\x$};
\end{tikzpicture}


%% file: figs/labelex.tex



\begin{tikzpicture}[scale=3,>=stealth]
	\tikzstyle{class1}=[circle,draw,fill=red,inner sep=0pt, minimum size = 1mm]
	\tikzstyle{class2}=[diamond,draw,fill=blue,inner sep=0pt, minimum size = 1mm]
	\tikzstyle{nonclass}=[circle,draw,inner sep=0pt, minimum size = 1mm]
	\tikzstyle{edge}=[rectangle,draw,fill=black,inner sep=0pt, minimum size = 1mm]
	\tikzstyle{tol}=[circle,draw,dotted,thick,inner sep=0pt, minimum size = 7mm]
	
	\draw[black] plot[domain=-1:1](\x,\x);

	\node at (-0.7,-0.8) [class1]{};
	\node at (-0.7,-0.6) [class2]{};
	\node at (-0.7,-0.7) [edge]{};
	\node at (-0.7,-0.7) [tol]{};

	
	\draw[<->] (-0.6,-0.3) -- (-0.5,-0.3)
		node[above=2pt,midway] {$\delta$};
	
	\node at (-0.5,-0.5) [edge]{};
	\node at (-0.5,-0.5) [tol]{};
	\node at (-0.4,-0.5) [class1]{};
	\node at (-0.6,-0.45) [class2]{};

	\node at (0,0.1) [edge]{};
	\node at (0,0.1) [tol]{};
	\node at (-0.1,0.1) [class2]{};
	\node at (0.1,0.1) [class1]{};

	\node at (0.55,0.5) [edge]{};
	\node at (0.55,0.5) [tol]{};
	\node at (0.55,0.4) [class1]{};
	\node at (0.45,0.58) [class2]{};

	\node at (0.2,0.6) [nonclass]{};
	\node at (0.7,-0.3)[nonclass]{};
	\node at (-0.5,0.3) [nonclass]{};
	\node at (-0.2,-0.8) [nonclass]{};

\end{tikzpicture}


%% file: figs/labelUS2.tex



\begin{tikzpicture}[scale=2,>=stealth]
	\tikzstyle{class1}=[circle,draw,fill=red,inner sep=0pt, minimum size = 1mm]
	\tikzstyle{class2}=[diamond,draw,fill=blue,inner sep=0pt, minimum size = 1mm]
	\tikzstyle{US}=[rectangle,draw,fill=green,inner sep=0pt, minimum size = 1mm]
	\tikzstyle{nonclass}=[circle,draw,inner sep=0pt, minimum size = 1mm]
	\tikzstyle{edge}=[rectangle,draw,fill=black,inner sep=0pt, minimum size = 1mm]
	\tikzstyle{tol}=[circle,draw,dotted,thick,inner sep=0pt, minimum size = 12mm]
	\tikzstyle{LIP1}=[circle,draw,dotted,thick,inner sep=0pt, minimum size = 20mm]
	\tikzstyle{LIP2}=[circle,draw,gray,solid,thick,inner sep=0pt, minimum size = 17mm]
	
	\draw[black] plot[domain=-1:1](\x,\x);

	\node at (-0.7,-0.8) [class1]{};
	\node at (-0.7,-0.6) [class2]{};
	
    \node at (-0.4,-0.5) [class1]{};
	\node at (-0.6,-0.4) [class2]{};

	\node at (-0.1,0.1) [class2]{};
	
    \node at (0.1,0.1) [class1]{};

	\node at (0.55,0.4) [class1]{};
    \node at (0.55,0.4) [LIP1]{};

	\node at (0.45,0.6) [class2]{};
    \node at (0.45,0.6) [LIP2]{};

	\node at (0.2,0.6) [nonclass]{};
	\node at (0.7,-0.3)[nonclass]{};
	\node at (-0.5,0.3) [nonclass]{};
	\node at (-0.2,-0.8) [nonclass]{};

    \node at (0.3,0.4) [US]{};

    \draw[dotted,thick] (-0.5,1.0) -- (-0.9,1.0)
            node[left=5pt] {$C_1$};
    
    \draw[solid,gray, thick] (-0.5,0.8) -- (-0.9,0.8)
            node[left=5pt] {$C_2$};
	
    \node at (-1.0,0) {Class 2};
    \node at (0.8,-0.8) {Class 1};
\end{tikzpicture}


%% file: results.tex
\section{Numerical examples}\label{sec:results}

We now demonstrate the performance of the new discontinuity detection algorithm on a variety of problems: separating surfaces of varying complexity, a problem where the jump size varies along the discontinuity, and discontinuities of increasing dimension. Then we apply the algorithm to an ODE system whose fixed point depends discontinuously on its parameters, and finally we evaluate the performance of the algorithm on a problem where a discontinuity exists in a subspace of the full parameter domain.

\subsection{Geometry of the separating surface}\label{sec:diffreg}

\begin{figure}
\begin{center}
\subfigure[Initial classifier and labeled points from PA.]{\includegraphics[scale=0.32]{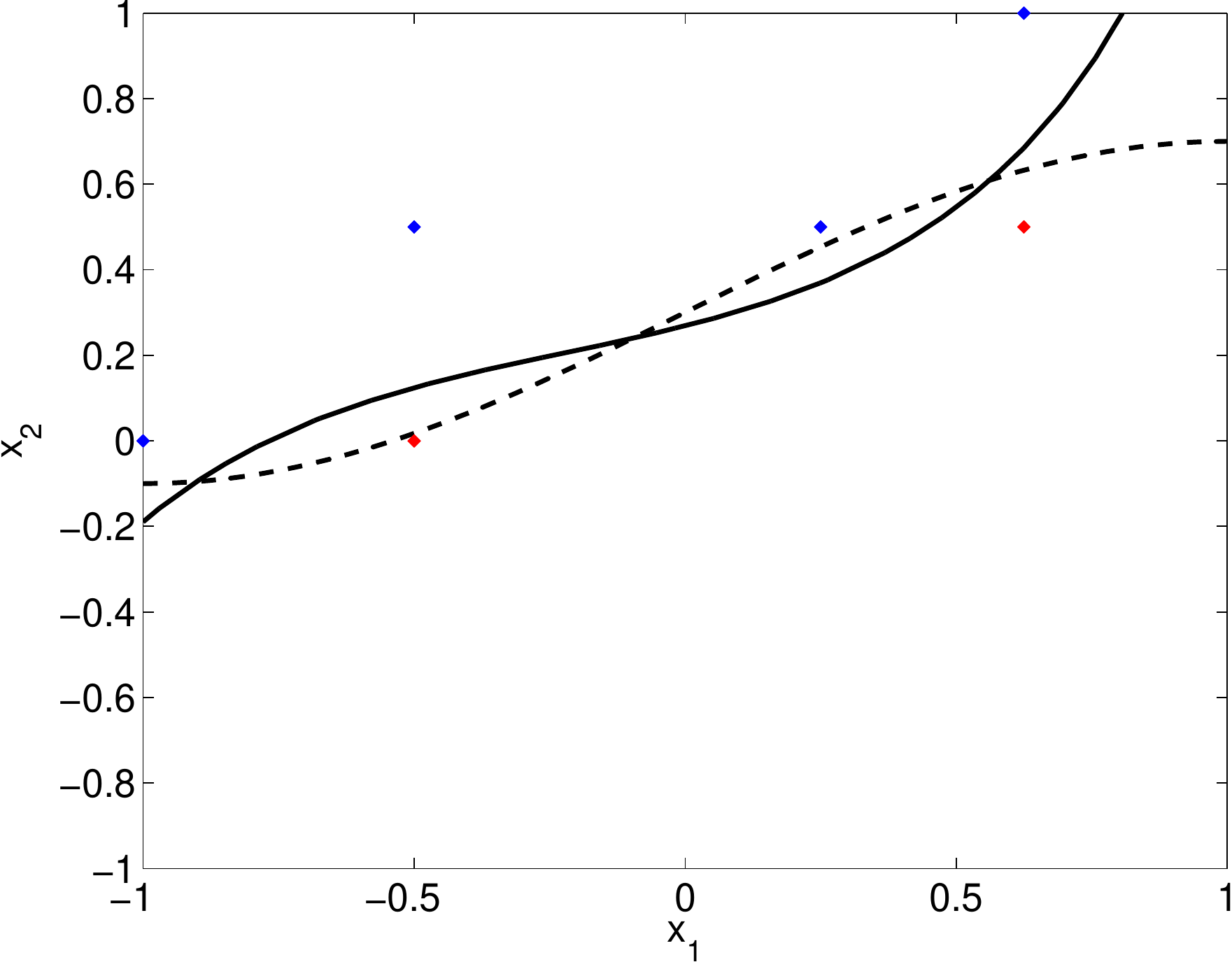}} 
\subfigure[Results after 5 iterations.]{\includegraphics[scale=0.32]{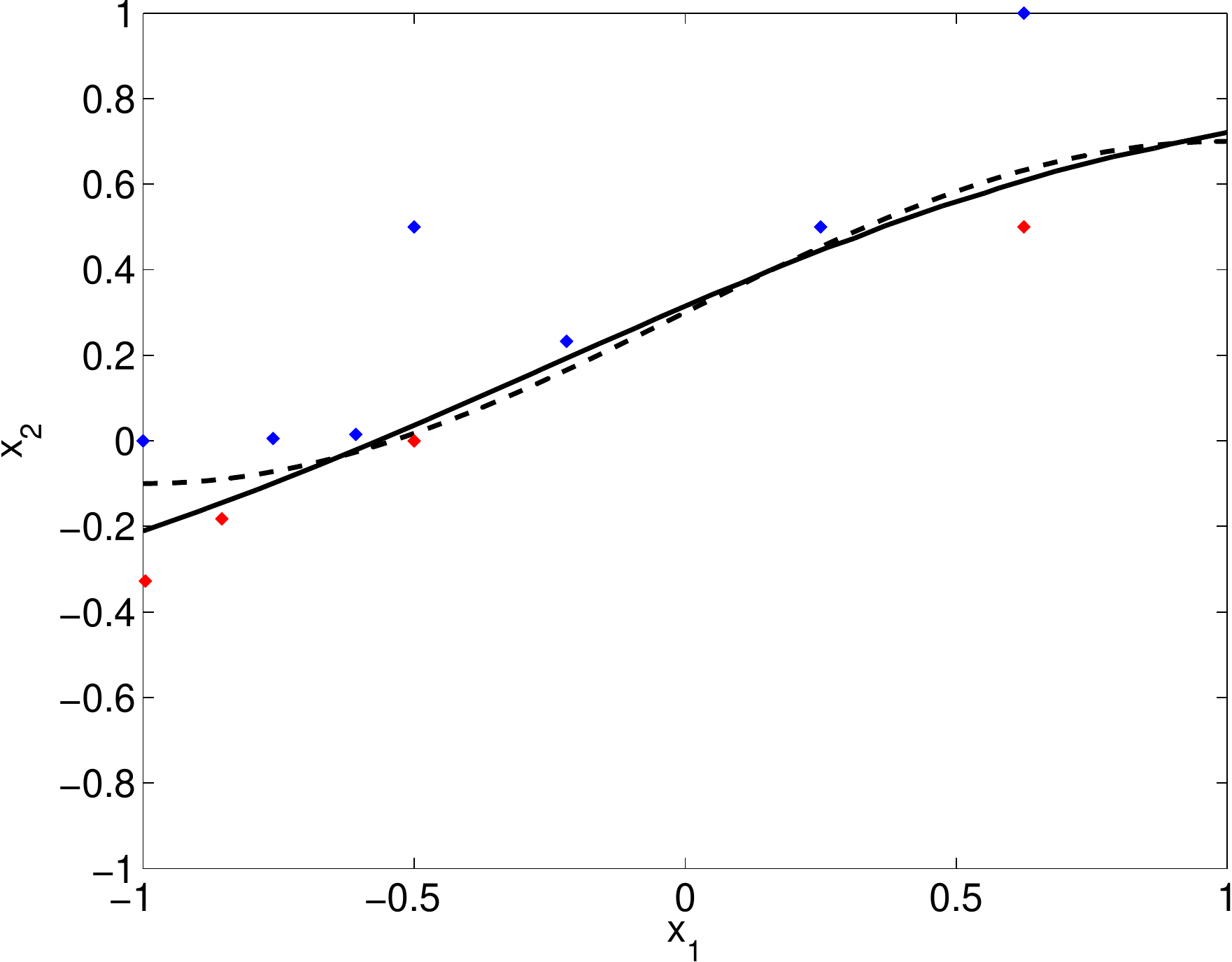}} \\
\subfigure[Results after 10 iterations.]{\includegraphics[scale=0.32]{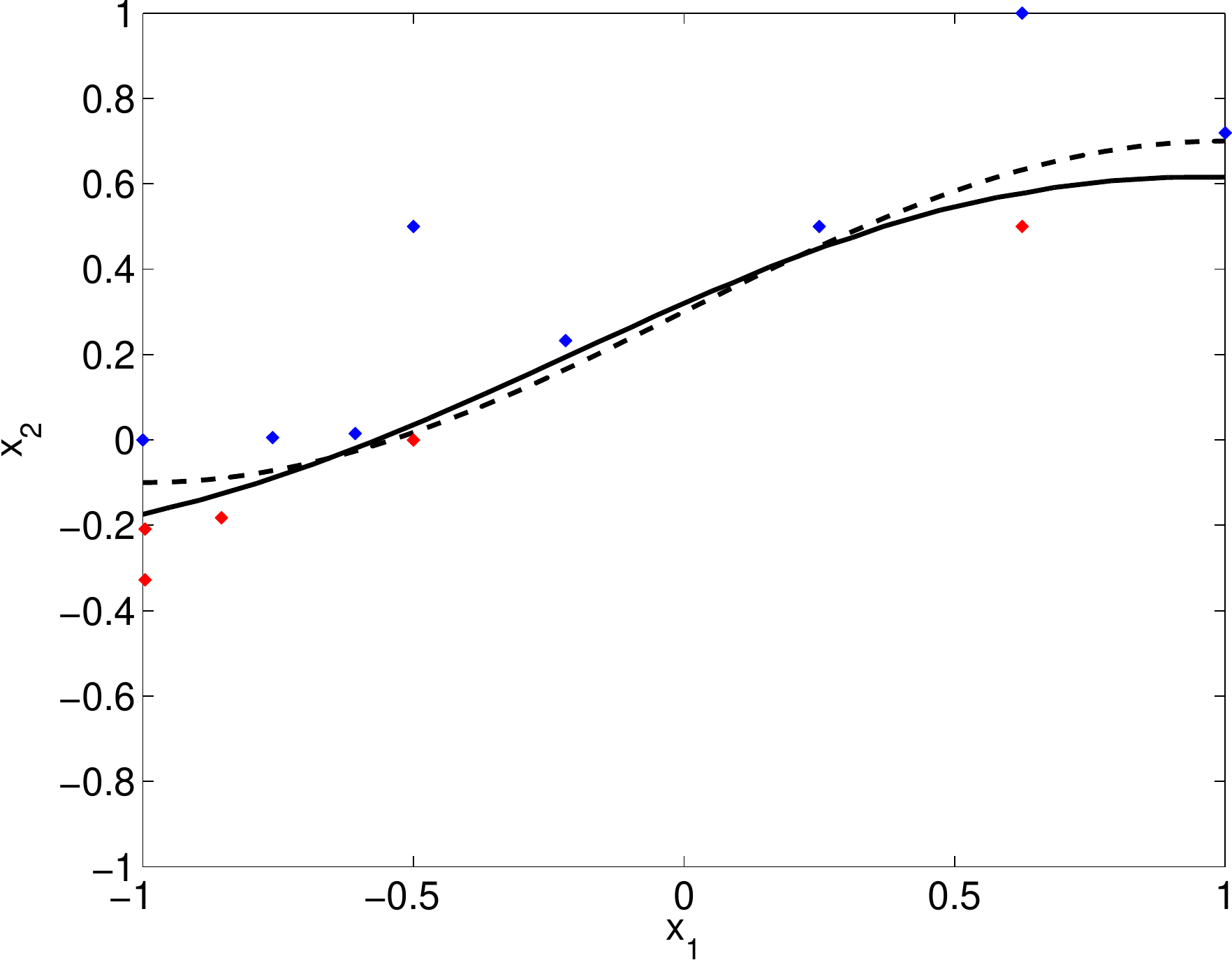}}
\subfigure[Results after 15 iterations.]{\includegraphics[scale=0.32]{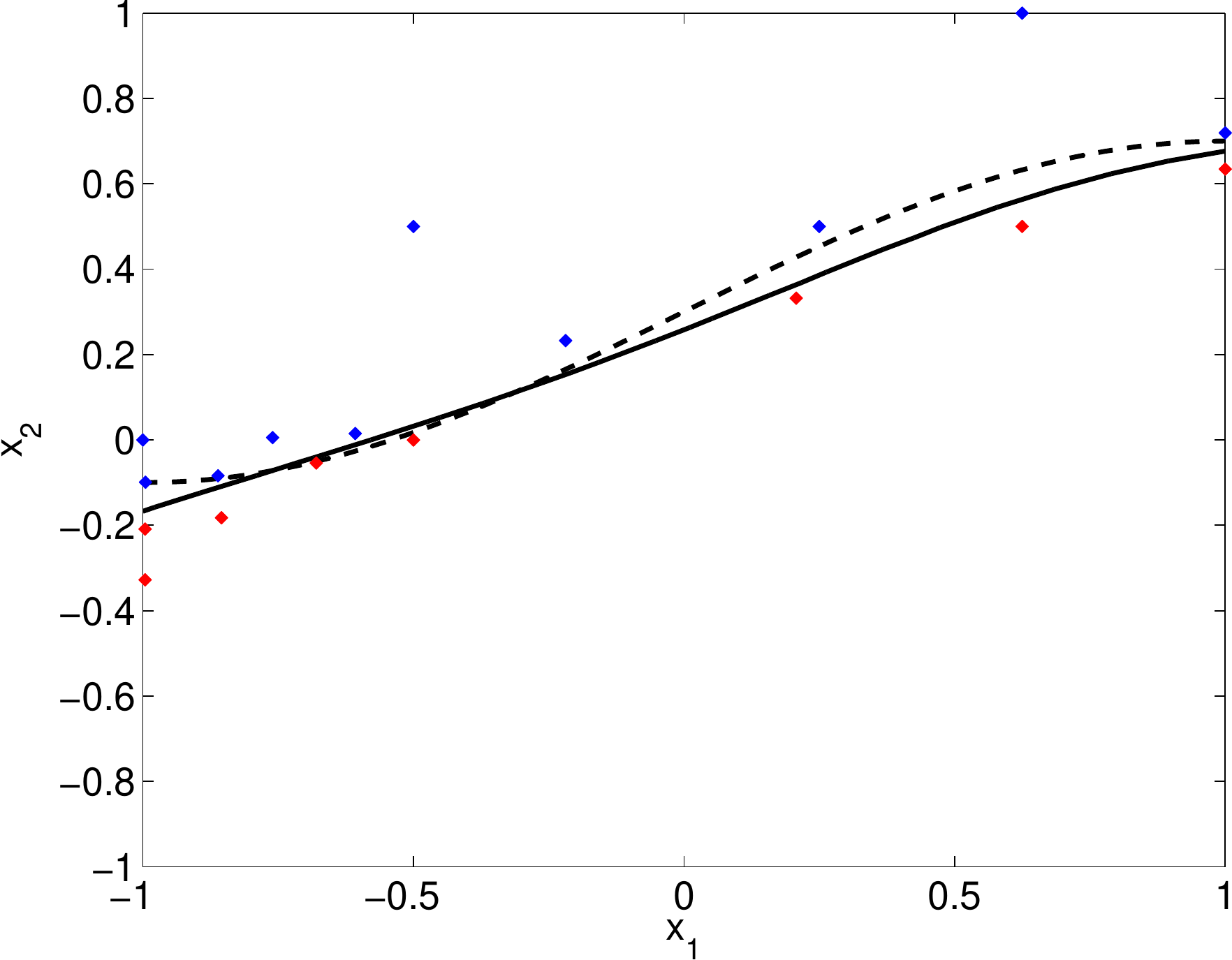}}
\caption{Uncertainty sampling results for separating surface given in Equation~\ref{eq:surf1}. Red and blue points are training samples from each class. The dotted line represents the true separating surface, the solid black line represents the zero level set of the SVM.}
\label{fig:USapplied1}
\end{center}
\end{figure}

\begin{figure}
\begin{center}
\subfigure[Initial classifier and labeled points from PA.]{\includegraphics[scale=0.3]{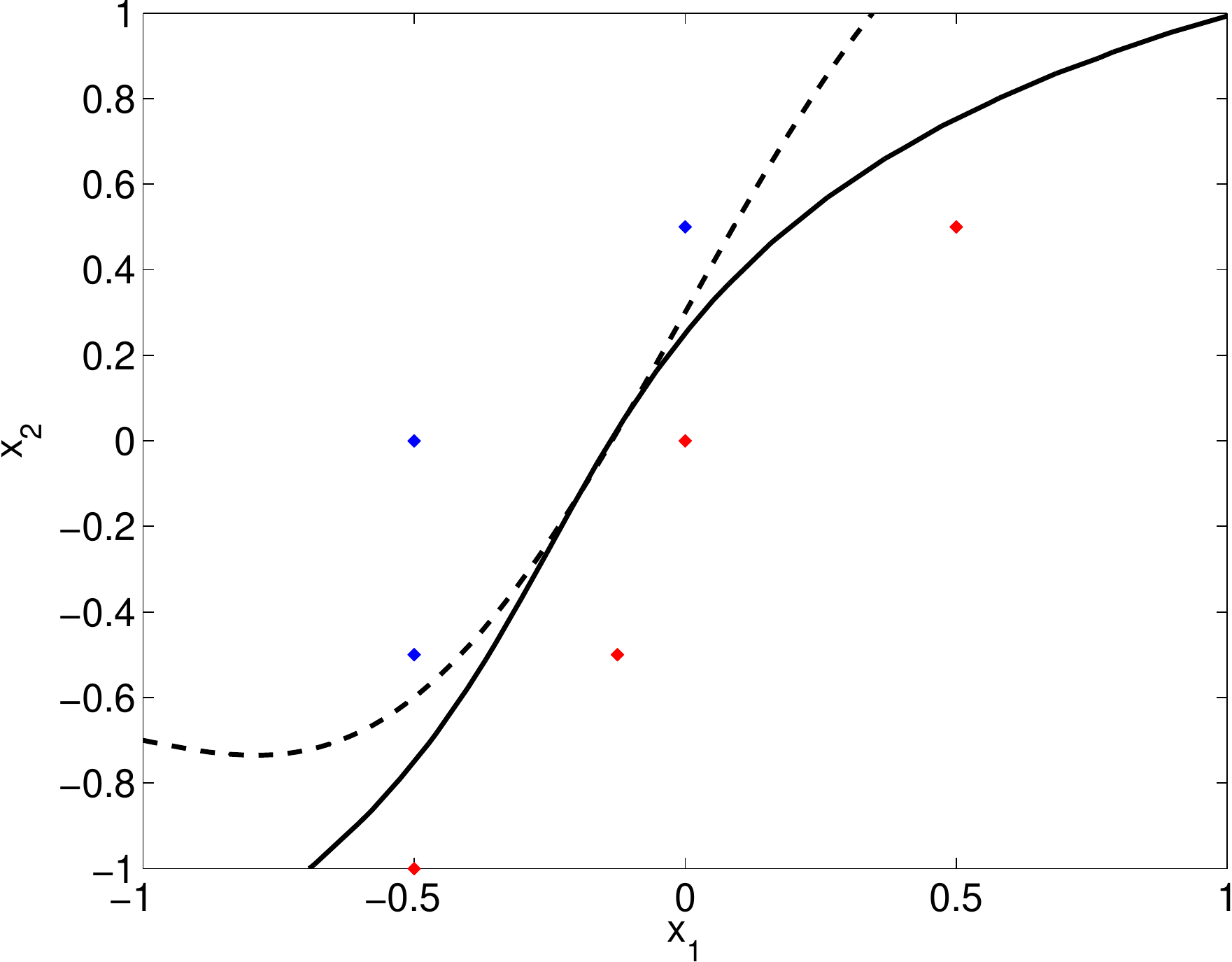}} 
\subfigure[Results after 10 iterations.]{\includegraphics[scale=0.3]{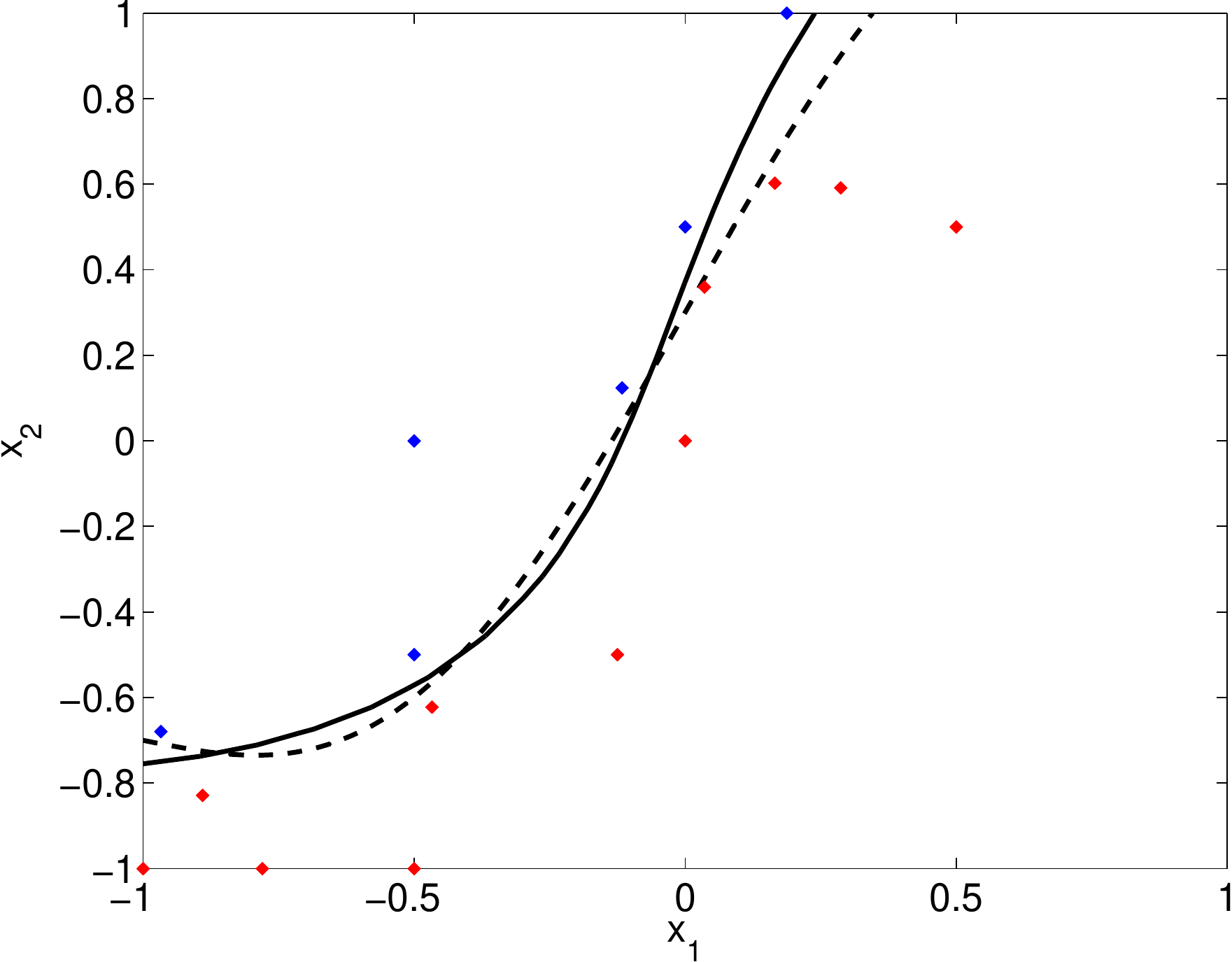}} \\
\subfigure[Results after 15 iterations.]{\includegraphics[scale=0.3]{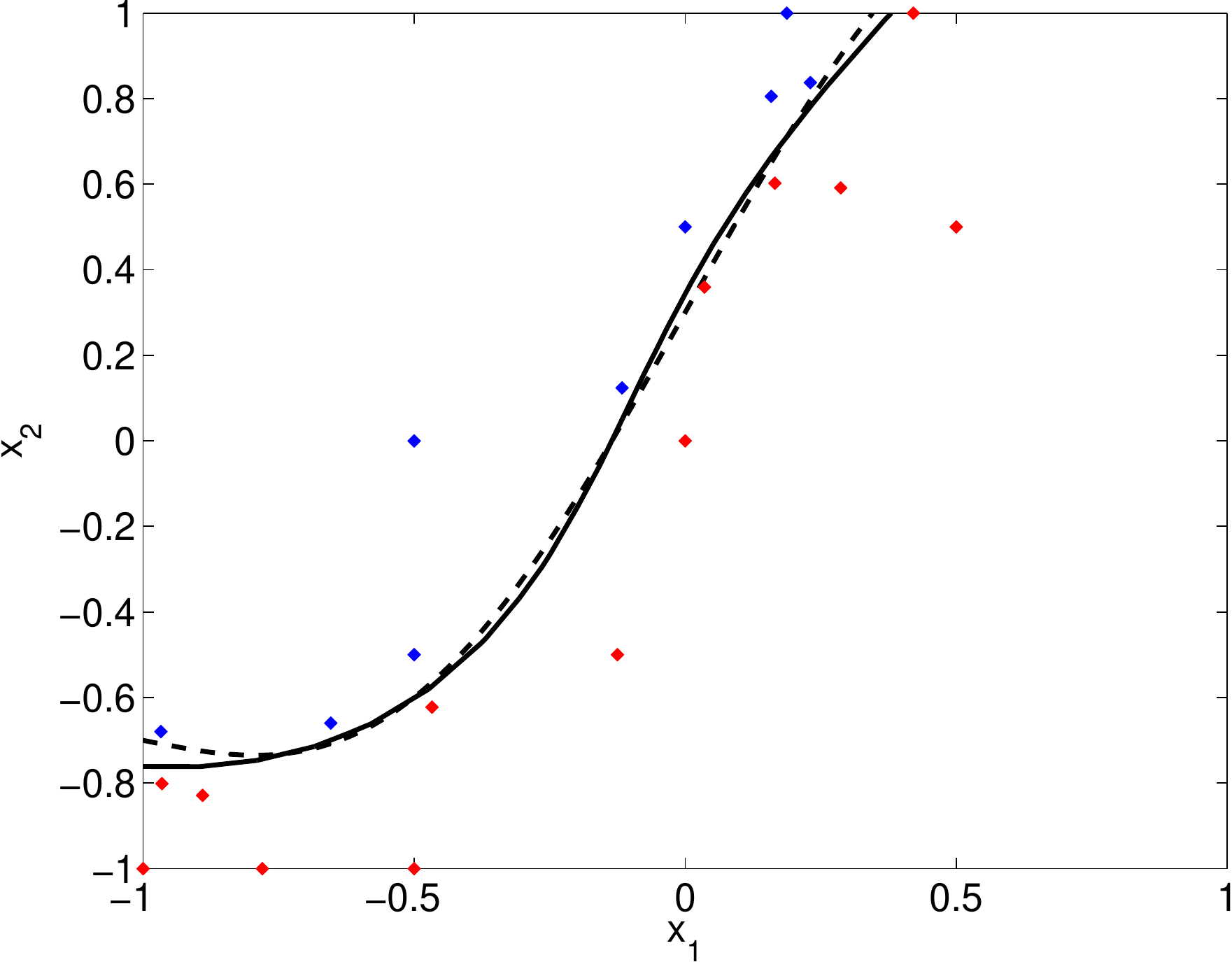}} 
\subfigure[Results after 20 iterations.]{\includegraphics[scale=0.3]{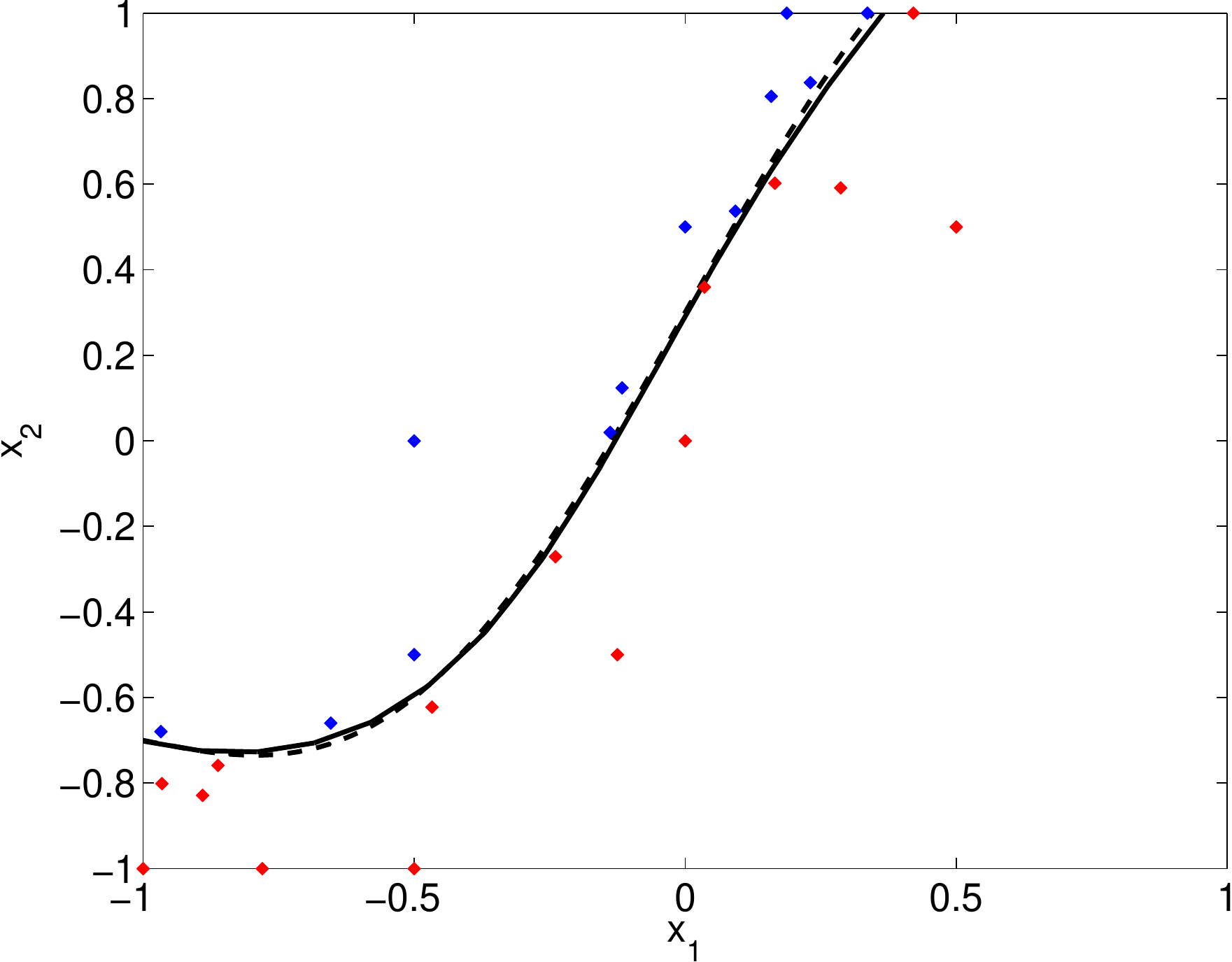}} 
\caption{Uncertainty sampling results for separating surface given in Equation~\ref{eq:surf2}. Red and blue points are training samples from each class. The dotted line represents the true separating surface, the solid black line represents the zero level set of the SVM.}
\label{fig:USapplied2}
\end{center}
\end{figure}

\begin{figure}
\begin{center}
\subfigure[Initial classifier and labeled points from PA.]{\includegraphics[scale=0.32]{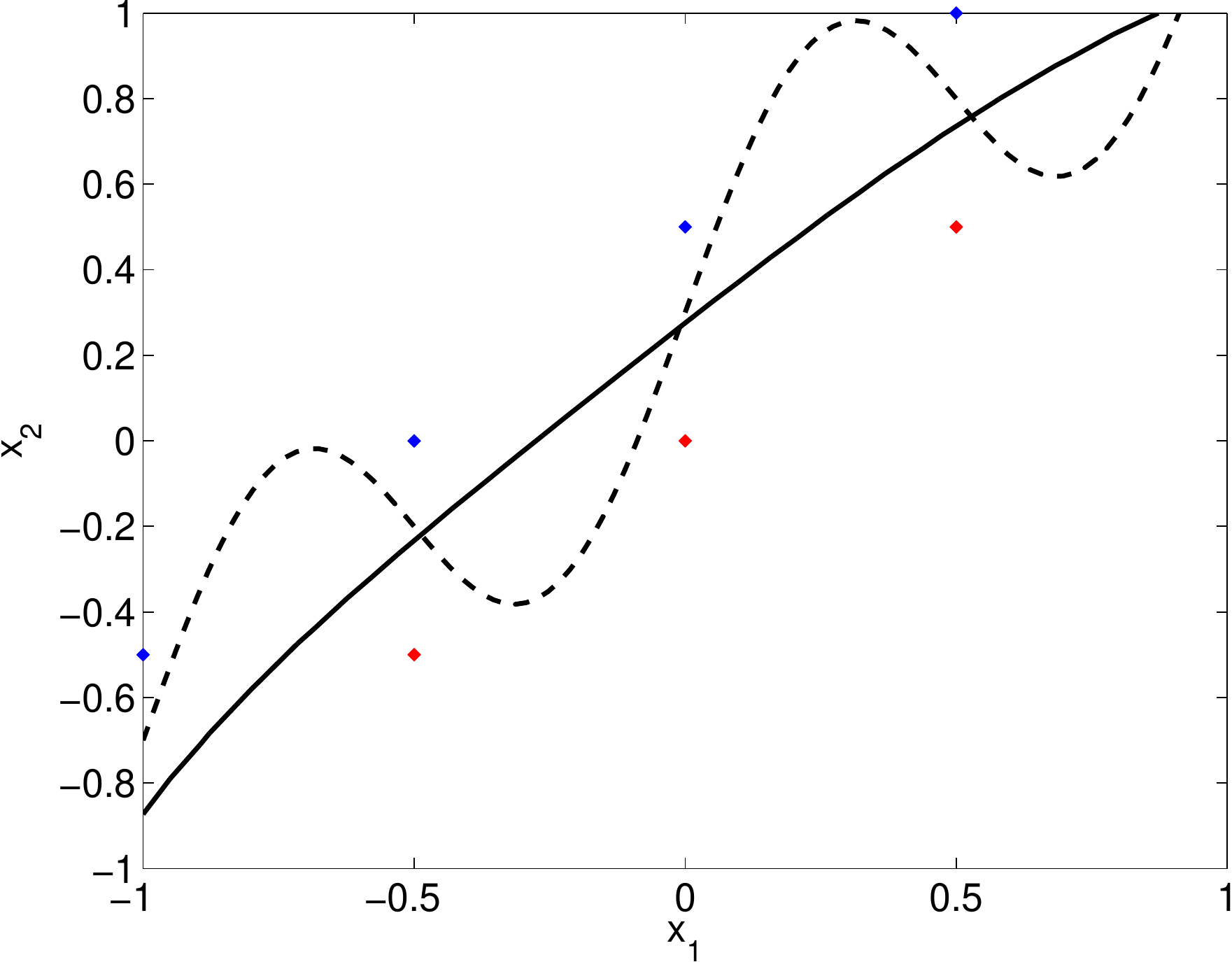}}
\subfigure[Results after 25 iterations.]{\includegraphics[scale=0.32]{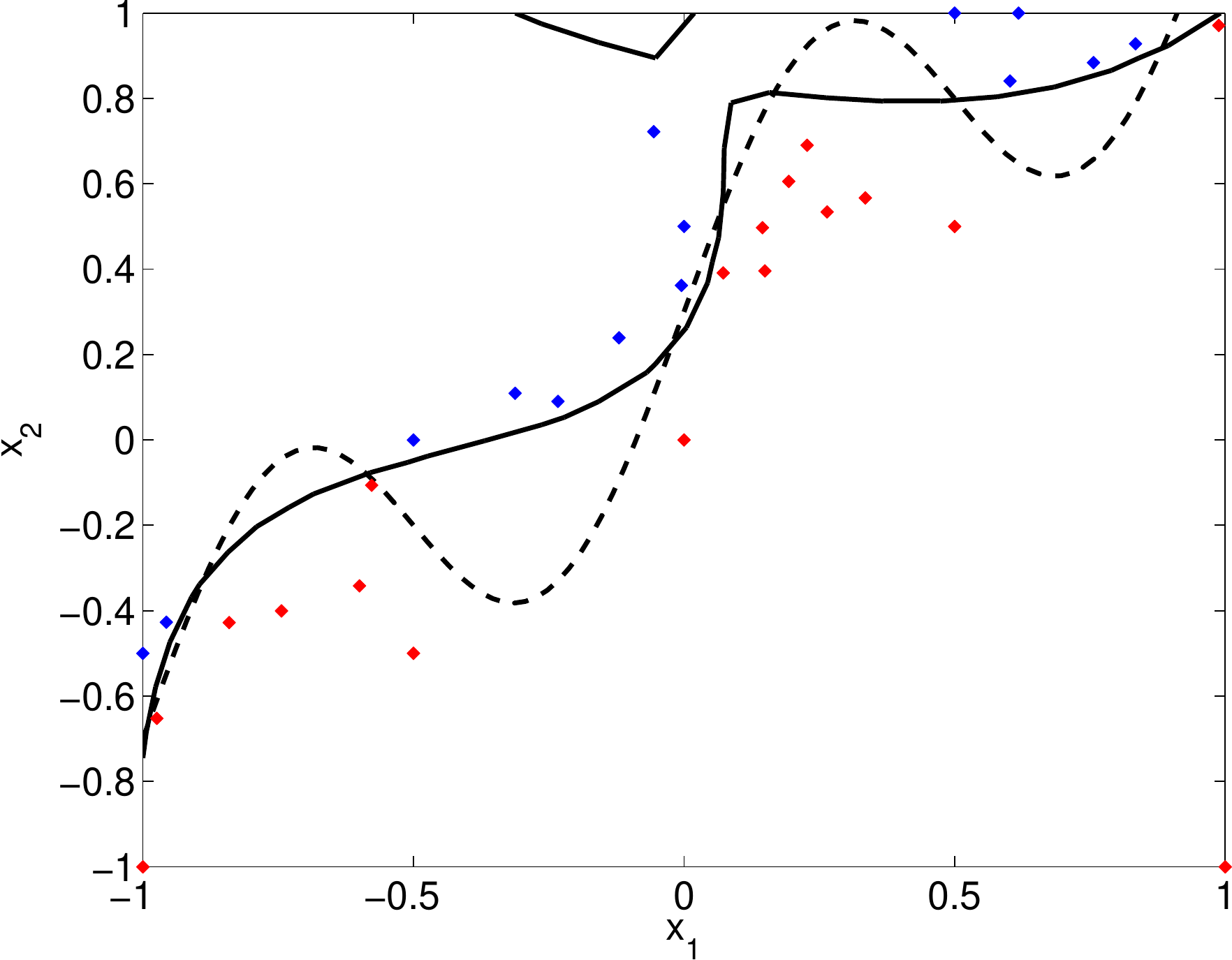}} \\
\subfigure[Results after after 50 iterations.]{\includegraphics[scale=0.32]{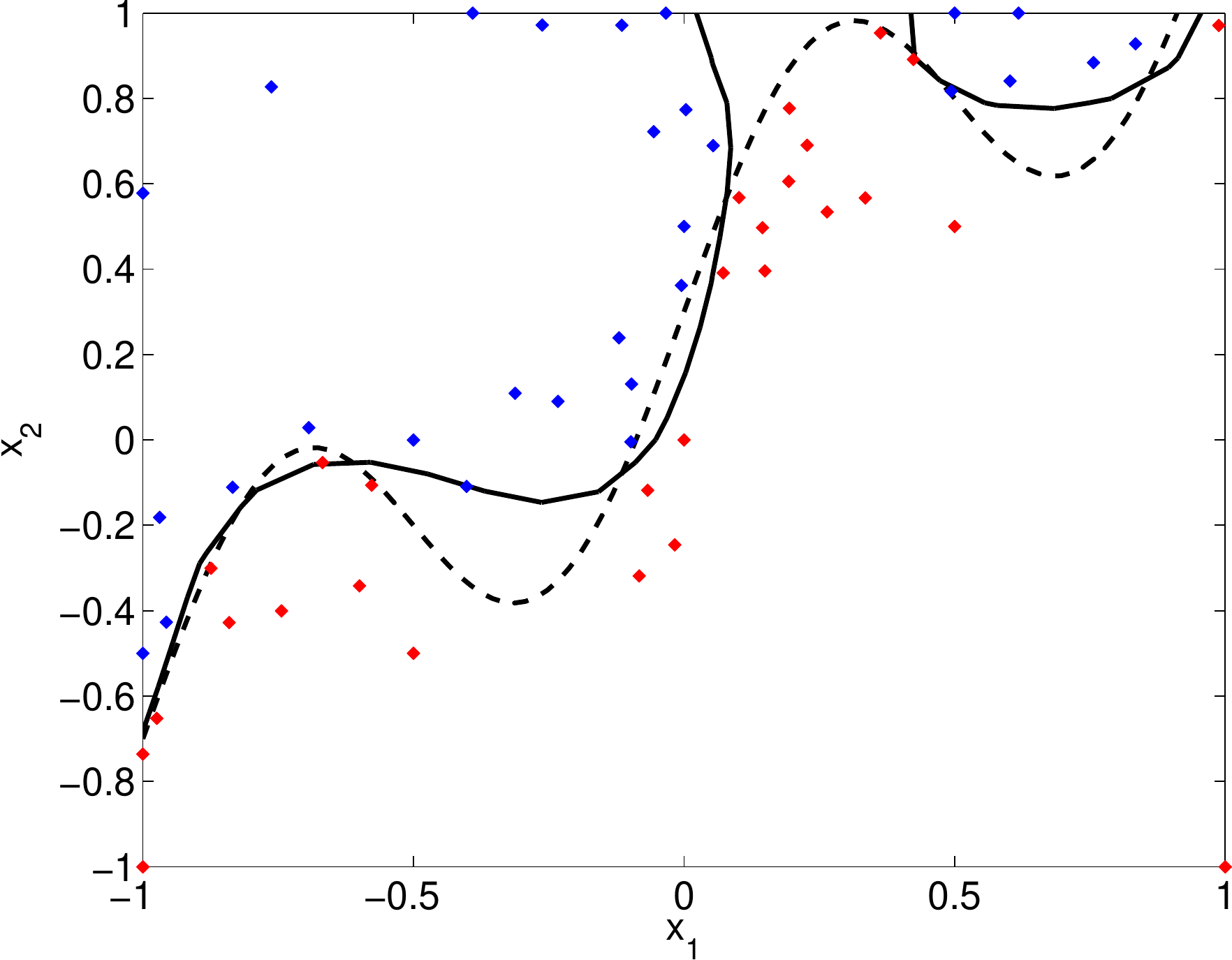}}
\subfigure[Results after 75 iterations.]{\includegraphics[scale=0.32]{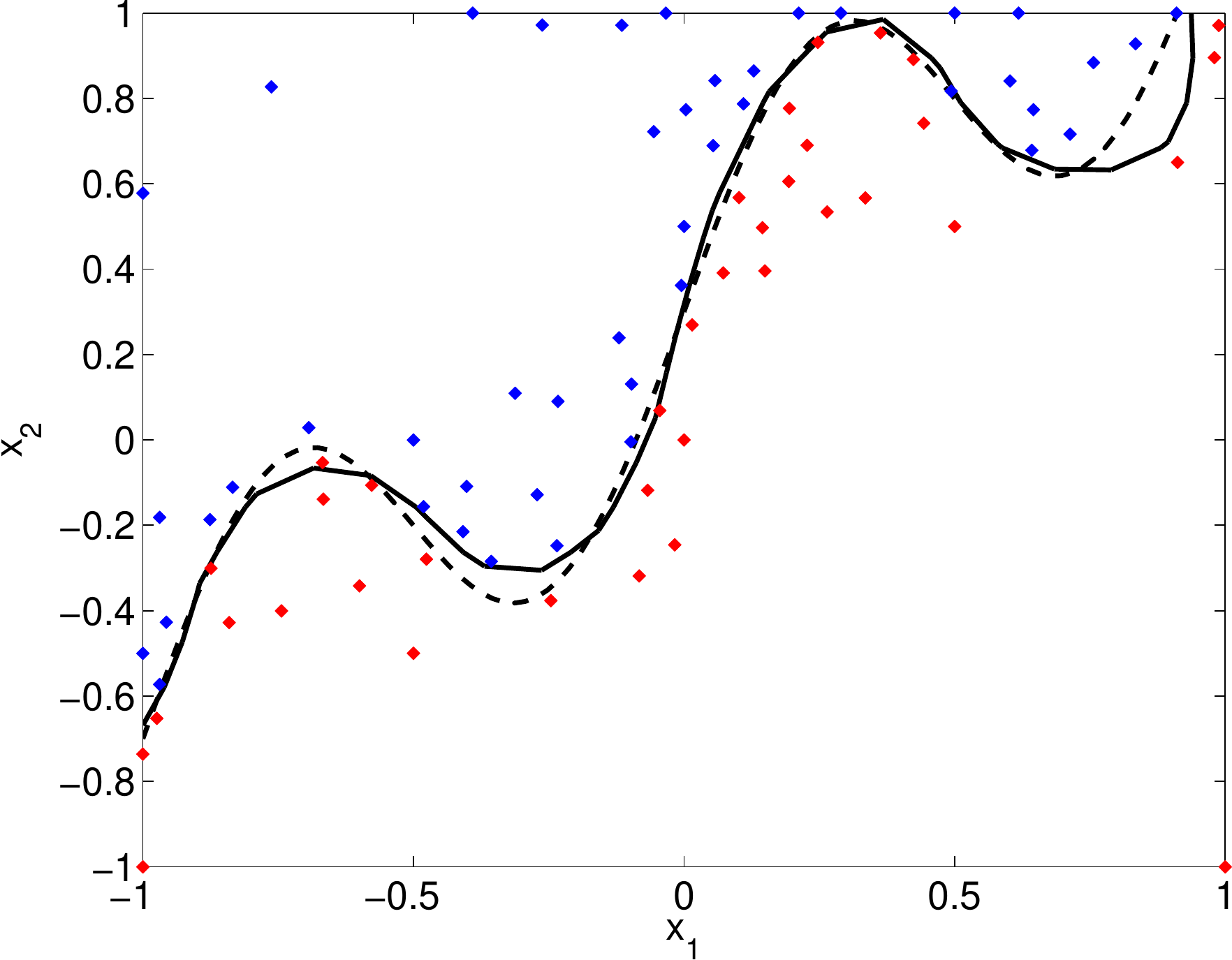}}
\caption{Uncertainty sampling results for separating surface given in Equation~\ref{eq:surf3}. Red and blue points are training samples from each class. The dotted line represents the true separating surface, the solid black line represents the zero level set of the SVM.}
\label{fig:USapplied3}
\end{center}
\end{figure}

\begin{figure}
\begin{center}
\subfigure[Initial classifier and labeled points obtained from PA.]{\includegraphics[scale=0.32]{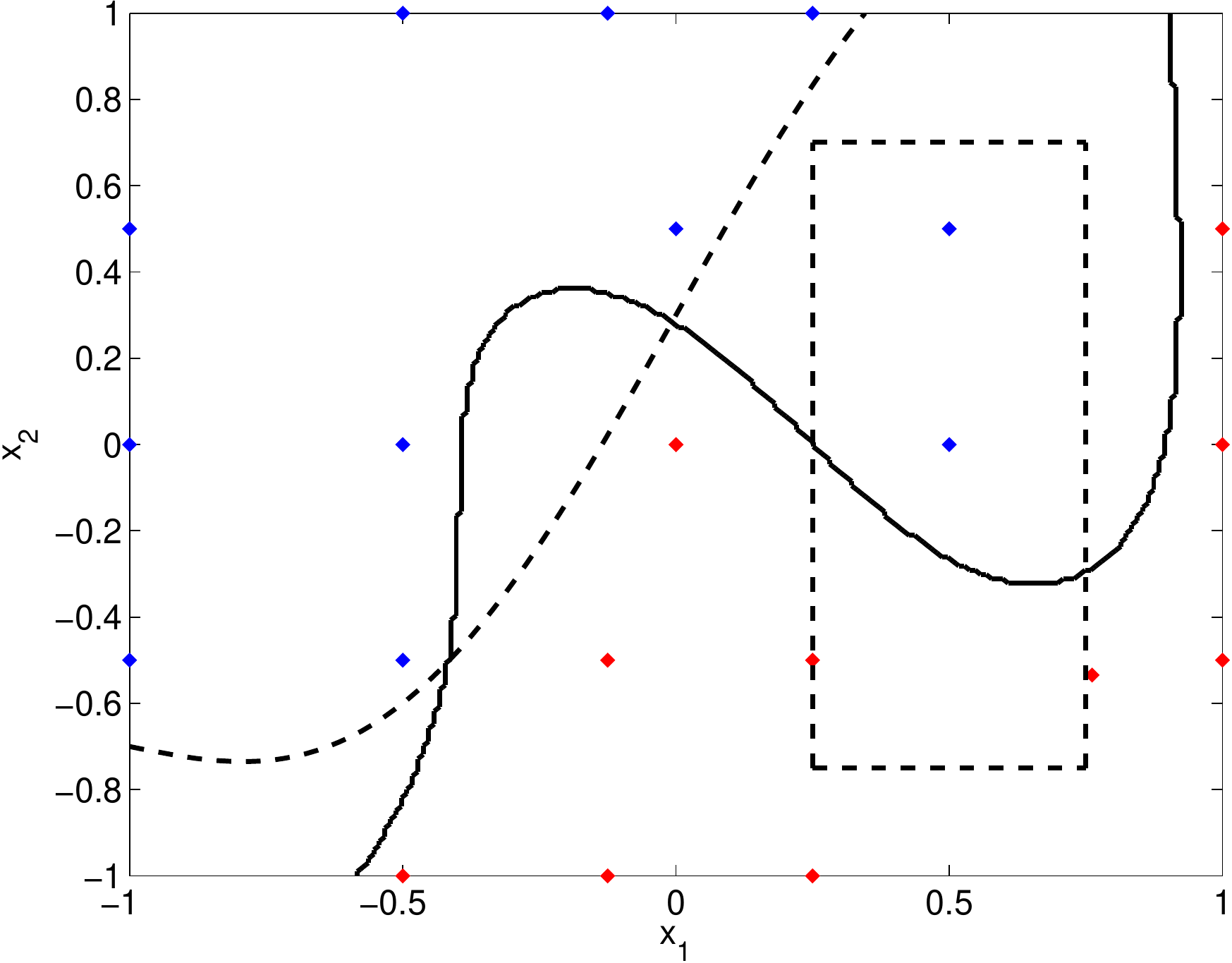}}
\subfigure[Results after 50 iterations.]{\includegraphics[scale=0.32]{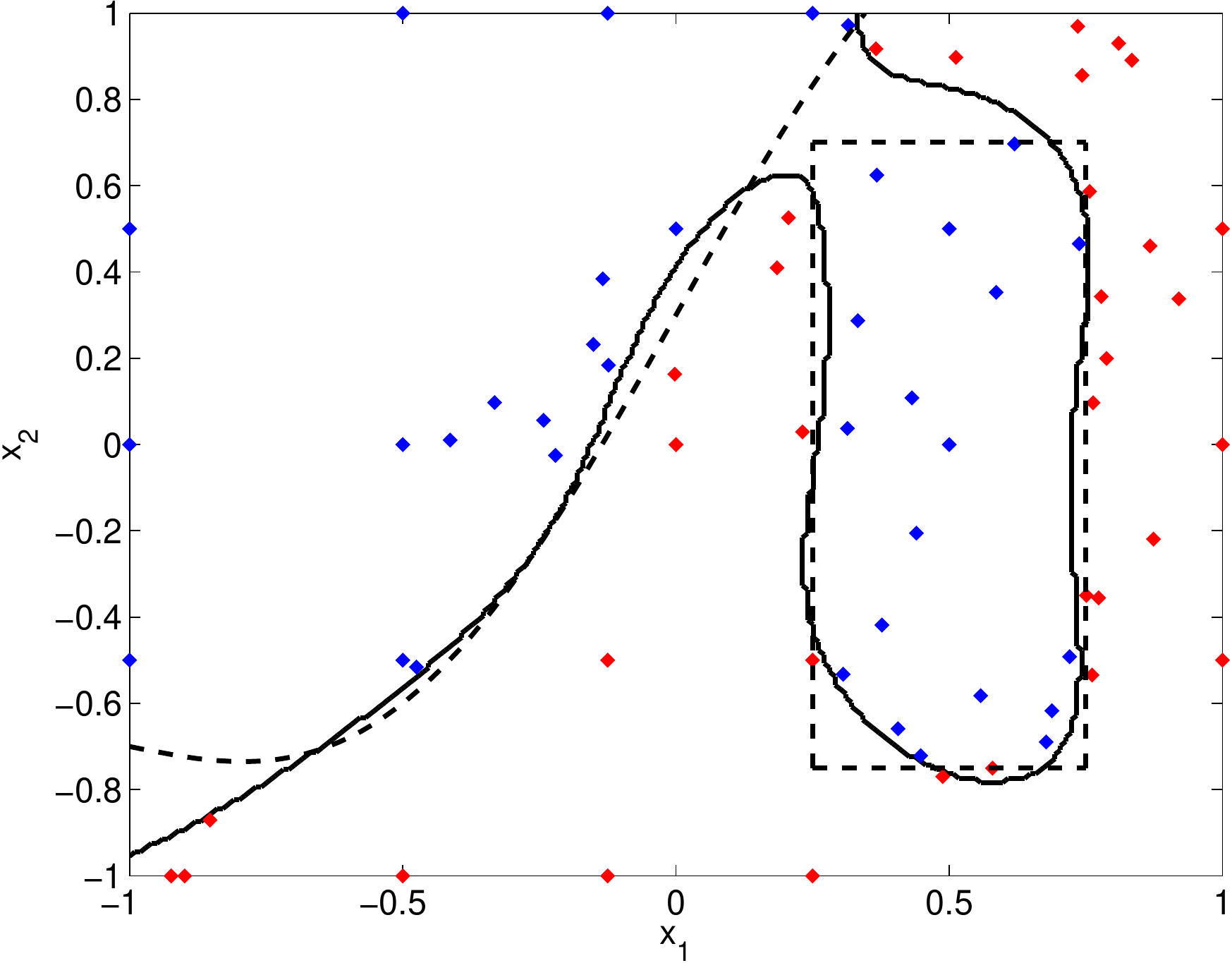}} \\
\subfigure[Results after after 100 iterations.]{\includegraphics[scale=0.32]{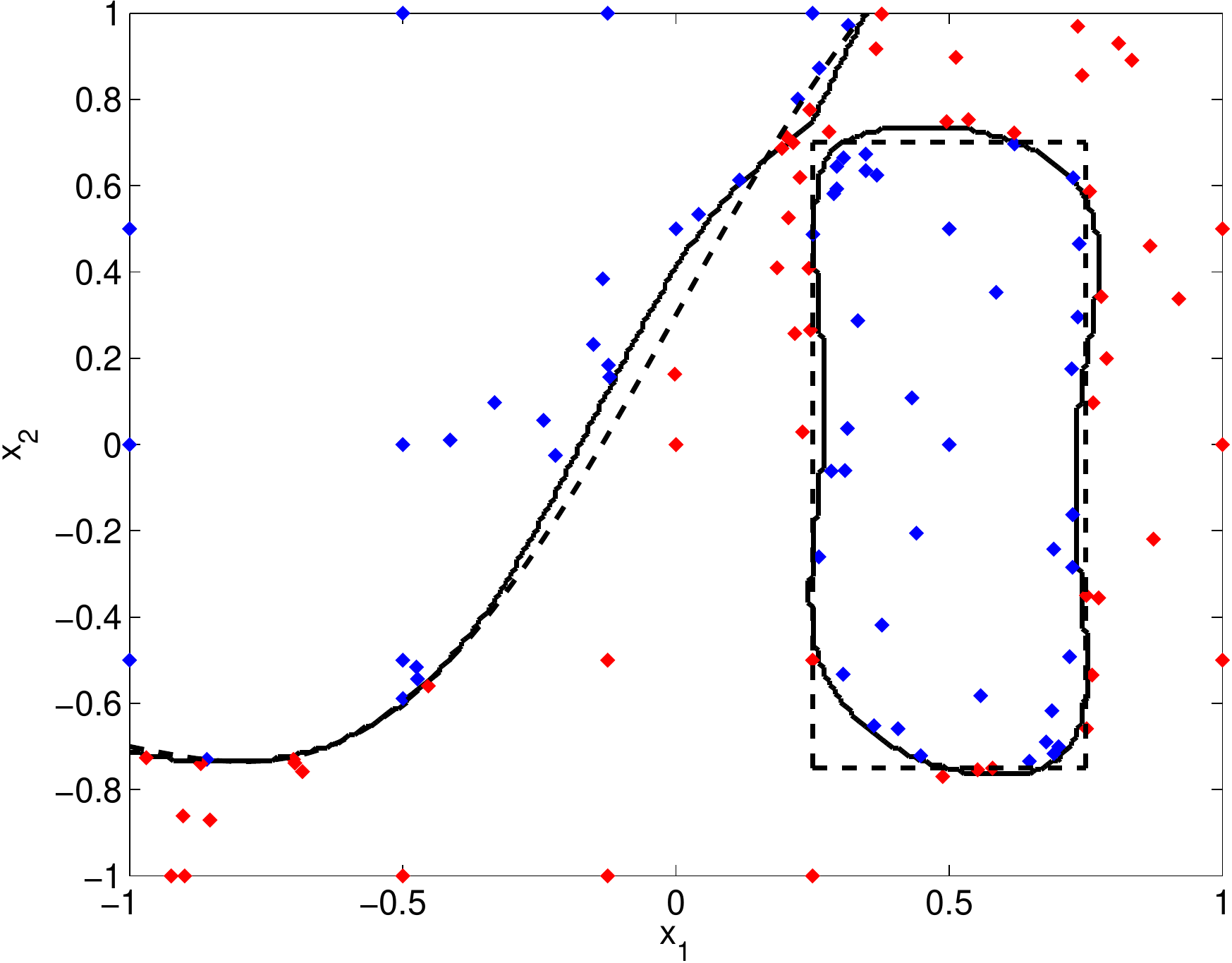}}
\subfigure[Results after 120 iterations.]{\includegraphics[scale=0.32]{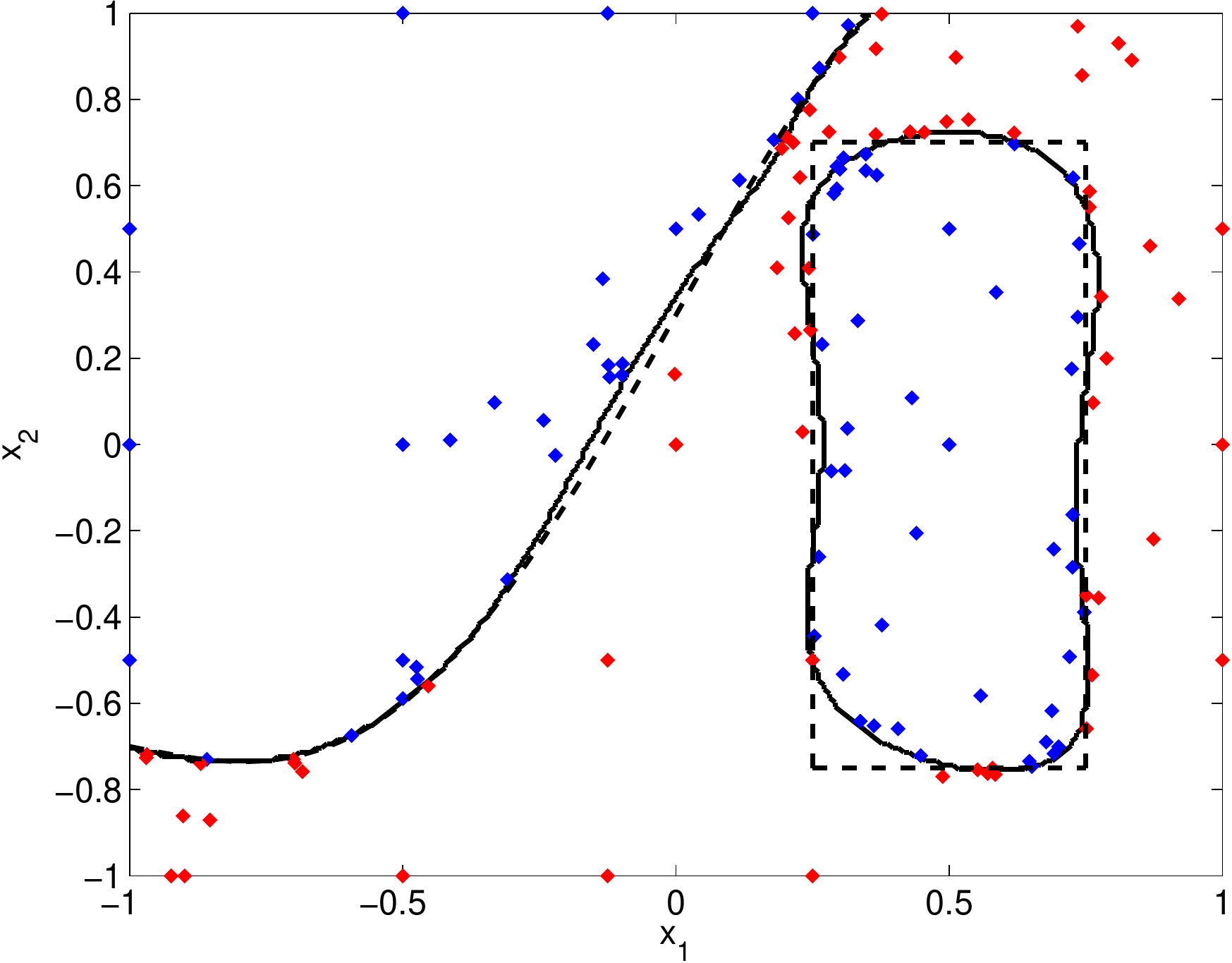}}
\caption{Uncertainty sampling results for separating surface given Equation~\ref{eq:surf2} with the addition of a box. Red and blue points are training samples from each class. The dotted line represents the true separating surface, the solid black line represents the zero level set of the SVM.}
\label{fig:USapplied4}
\end{center}
\end{figure}

To evaluate how the performance of the algorithm depends on the regularity of the separating surface, we consider four increasingly complex discontinuity geometries, all in the two-dimensional parameter space $D = [-1,1]^2$. The first three separating surfaces are given by (\ref{eq:surf1})--(\ref{eq:surf3}) and illustrated in Figures~\ref{fig:USapplied1}--\ref{fig:USapplied3}. The final separating surface is a combination of (\ref{eq:surf2}) and a rectangle, and serves as an example of a discontinuity bounding regions that are not simply connected; this surface is illustrated in Figure~\ref{fig:USapplied4}.
\begin{eqnarray}
x_2 & = & 0.3+0.4\sin(\pi x_1) 
\label{eq:surf1} \\
x_2 & = & 0.3+0.4\sin(\pi x_1) + x_1
\label{eq:surf2} \\
x_2 & = & 0.3+0.4\sin(2\pi x_1)  + x_1 
\label{eq:surf3}
\end{eqnarray}
Because this example is intended to focus on the geometry of the separating surface, the function in the two regions simply takes values of $+1$ and $-1$. 

The initialization phase of the algorithm is performed with an equal off-axis and edge tolerance $\mathtt{tol} = \delta =0.5$. Uncertainty sampling is performed with $\delta_t=2$ and $\epsilon=0.01$, indicating that we believe the variation of the function is small compared to the jump size and allowing newly sampled points to be fairly close together. We note that in these scenarios the function is constant (exhibits no variation) within each class. The SVM classifier is trained using Gaussian kernels, with parameters chosen via cross-validation.  Figures \ref{fig:USapplied1}--\ref{fig:USapplied4} show how the distributions of positively/negatively labeled points and the classifier boundary evolve after various iterations of uncertain sampling. The first discontinuity (\ref{eq:surf1}) is almost linear and requires the smallest number of function evaluations to be accurately captured. The second discontinuity (\ref{eq:surf2}) is fairly linear over a large region but contains a tail near the lower left hand corner. The refinement phase of the algorithm effectively locates this tail and accurately creates an approximation of the separating surface. The third discontinuity (\ref{eq:surf3}) has an oscillatory separating surface and requires the largest number of function evaluations in order to create an accurate classifier. Results for the fourth discontinuity show that the uncertainty sampling/SVM approach is capable of identifying and refining separating surfaces that are disjoint.

\begin{figure}
\begin{center}
\includegraphics[scale=0.5]{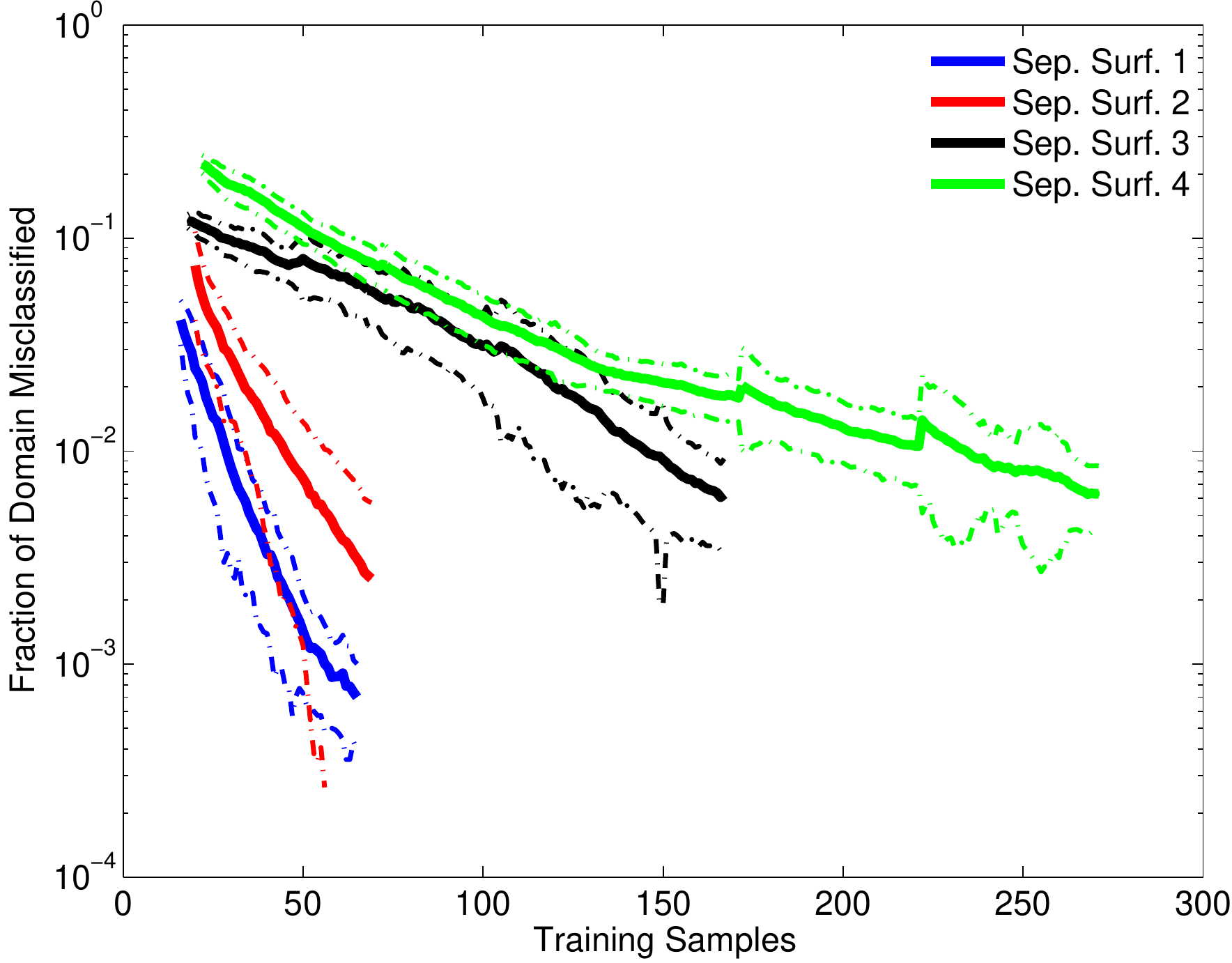}
\caption{Classifier convergence for different discontinuity geometries, specified in Section~\ref{sec:diffreg}. The fraction of the parameter domain that is misclassified is plotted versus the number of model evaluations. The discontinuity detection algorithm is run 100 times; dotted lines indicate errors $\pm$ one standard deviation away from the mean. The classifier for each separating surface achieves an error of less than 1\%.}
\label{fig:USconvergence}
\end{center}
\end{figure}

A quantitative assessment of classifier accuracy and convergence is given in Figure~\ref{fig:USconvergence}. To describe the classifier accuracy, we consider 10000 points sampled from a uniform distribution on the parameter domain and evaluate the percentage of these points that are classified incorrectly. For each of the four discontinuities, we plot the fraction of misclassified points versus the number of model evaluations. Since the discontinuity detection algorithm involves random sampling, we actually run it 100 times for each case and plot the mean and standard deviation of the misclassification percentage. Clearly, the more complex discontinuity geometries require more points to achieve an accurate approximation. But errors below 1\% are achieved for all four cases.

\subsection{Variable jump size} \label{sec:burgers}

\begin{figure}
\begin{center}
\includegraphics[scale=0.5]{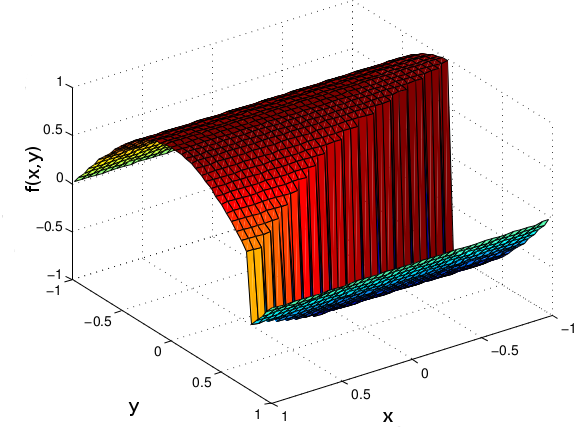}
\caption{Steady state solution of Burgers' equation plotted as a function of the spatial coordinate $x$ and the initial condition parameter $y$. The jump size varies along the discontinuity.}
\label{fig:burg}
\end{center}
\end{figure}

Now we demonstrate the performance of our algorithm on a discontinuity whose jump size is not constant along the separating surface. An example proposed in~\cite{Chantrasmi2009} of such a discontinuity is given in Figure~\ref{fig:burg}. This function arises from the solution of Burgers' equation with a parameterized initial condition. In particular, we have:
\begin{equation}
\frac{\partial u}{\partial t} + \frac{\partial}{\partial x}\left(\frac{u^2}{2}\right) = \frac{\partial}{\partial x}\left(\frac{\sin^2{x}}{2}\right),\quad 0 \leq x \leq \pi, \ t > 0,
\label{eq:burger}
\end{equation}
with initial condition $u(x,0) = y\sin{x}$, $y \sim \mathcal{U}(0,1)$, and boundary conditions $u(0,t) = u(\pi,t) = 0$. The surface in Figure~\ref{fig:burg} is the steady-state solution of (\ref{eq:burger}) plotted as a function of $y$, i.e., it is $\bar{u}(x, y) := u(x, t = \infty; y)$. 

In this experiment we set the labeling radius $\delta_t$ for uncertainty sampling points to $0.5$. A rationale for this choice is as follows. First, note that the minimum jump size in Figure~\ref{fig:burg}, occurring near the $(x,y)=(1,1)$ corner, is approximately $0.5$, with function values varying from $-0.25$ to $0.25$. Moving $0.5$ units away from this corner along the discontinuity, we find that function values near the discontinuity are approximately $f_{min}=-0.7$ and $f_{max}=0.7$, resulting in a jump size of $[f]=1.4$ and a half-jump size of $[f]/2 = 0.7$. Now suppose that one has already labeled some training points in this region and would like to label new uncertainty sampling points in the $(1,1)$ corner. If $f(x,y) \approx 0.25$ for some point $(x,y)$ in this corner, then $(x,y)$ is in $\textit{class 1}$ and $f_{max} - f(x,y) < [f]/2$; hence our labeling radius is valid for \textit{class 1}. Now consider the other class in the same corner. In the worst case situation, we will obtain a function value of $f(x,y) \approx -0.25$. Now $f(x,y) - f_{min} < [f]/2$, and therefore points from \textit{class 2} can be labeled as well.

\begin{figure}
\begin{center}
\includegraphics[scale=0.5]{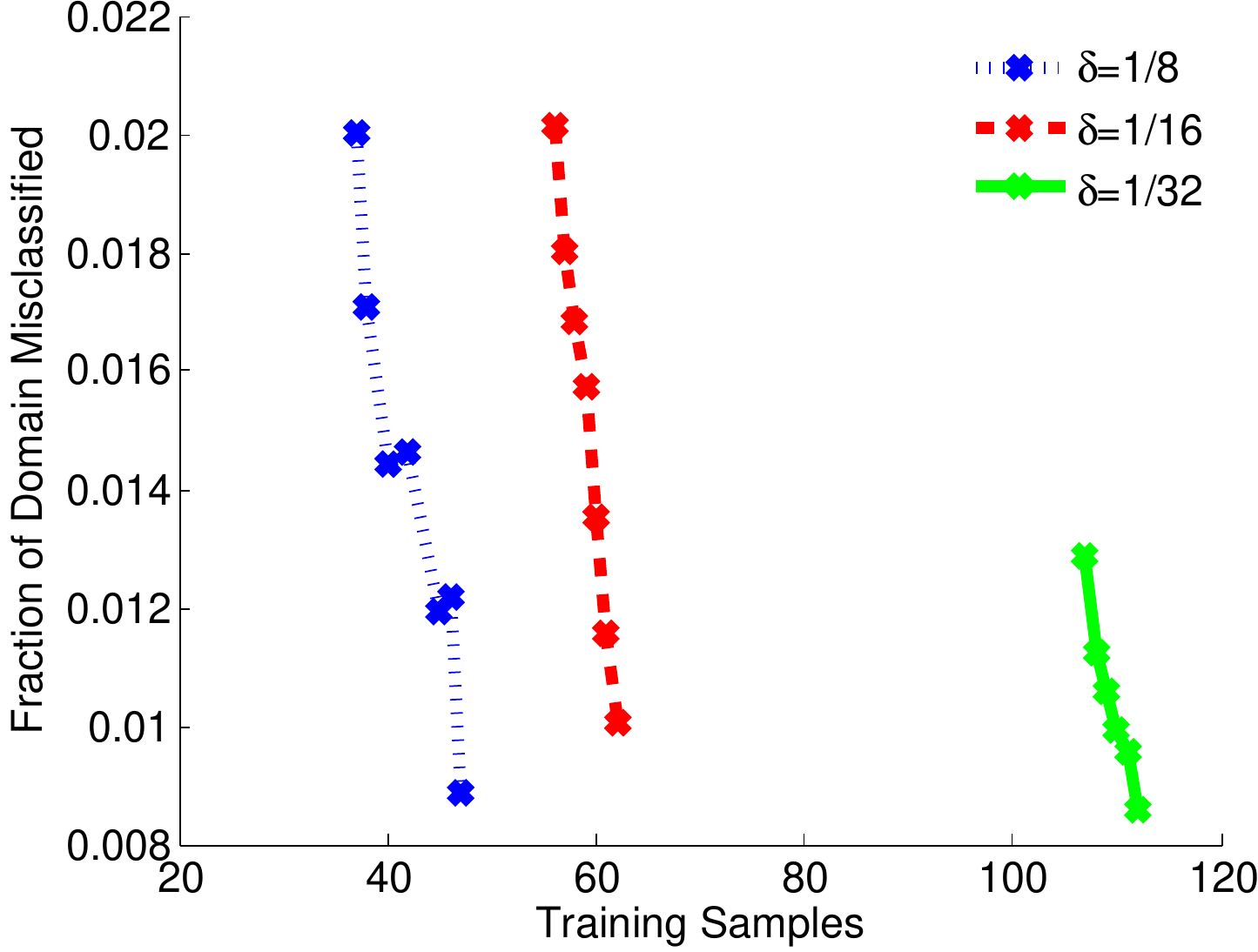}
\caption{Burgers' equation example: convergence of the discontinuity detection algorithm after initialization with different edge tolerances $\delta$.}
\label{fig:burgersConv}
\end{center}
\end{figure}

For this value of $\delta_t$, we now explore the impact of varying the edge tolerance $\delta$ in the polynomial annihilation phase of the algorithm. We consider $\delta \in \{ 1/8, 1/16, 1/32\}$ and show convergence results in Figure~\ref{fig:burgersConv}. For each of these refinement levels, we achieve a 1\% classification error on 5000 samples after approximately six uncertainty sampling iterations. Increasing the refinement of the polynomial annihilation phase of the algorithm increases the number of labeled samples initially fed to the SVM classifier, but this additional work does not seem to be useful. The active learning procedure correctly learns the discontinuity even when the initial PA grid is quite coarse.

\subsection{Dimension scaling} \label{sec:scaling}

\begin{figure}
\begin{center}
\includegraphics[scale=0.5]{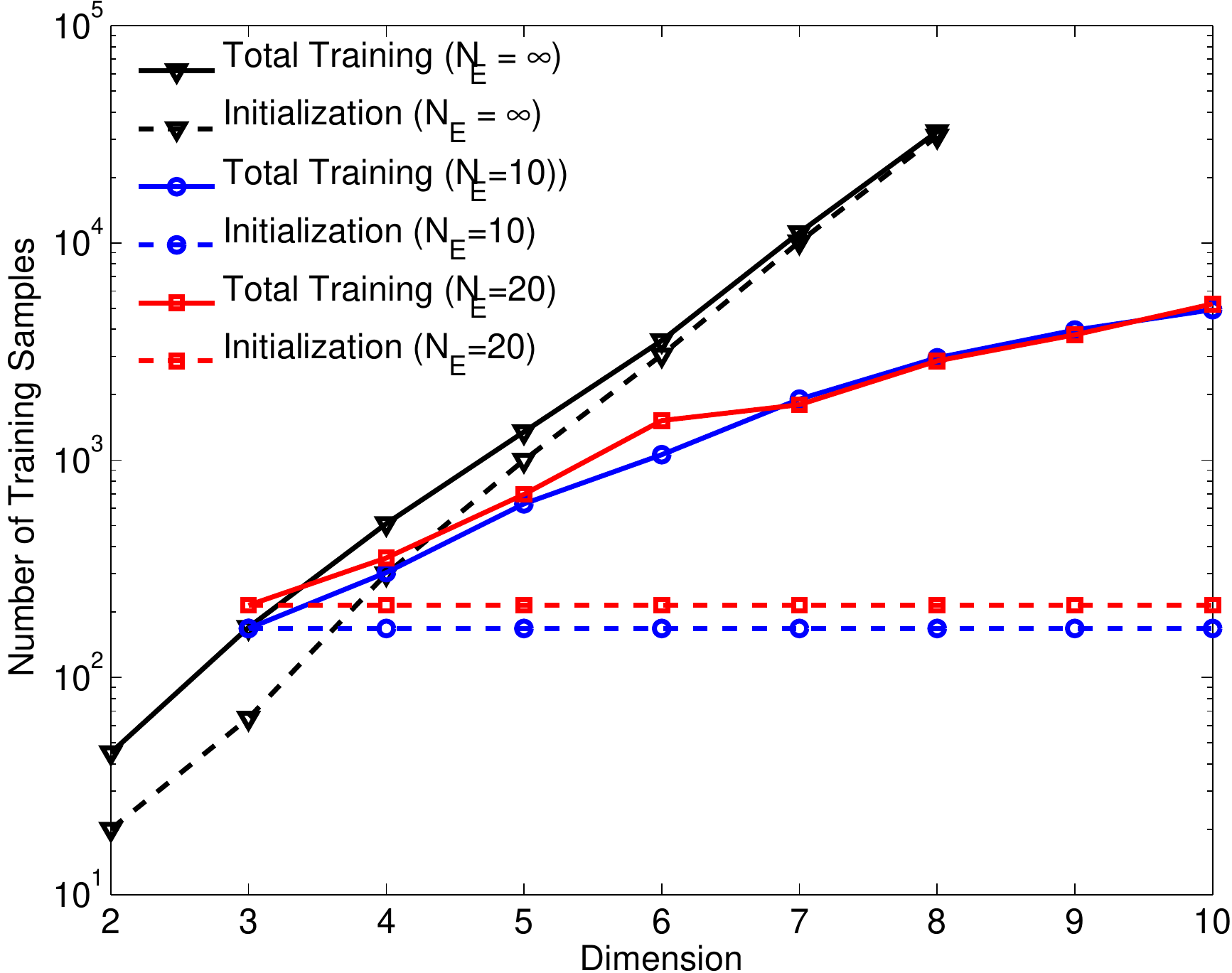}
\caption{Dimension scaling of different refinement algorithms for the cubic separating surface in Section~\ref{sec:scaling}. All algorithms are run until $99\%$ of the domain is correctly classified. Solid lines indicate the total number of function evaluations, while dashed lines indicate the number of function evaluations performed in the initialization phase of the algorithm only. Three cases are shown, corresponding to different limits $N_E$ on the number of edge points produced during initialization.}
\label{fig:scalePA}
\end{center}
\end{figure}

Now we evaluate the performance of the discontinuity detection algorithm on higher dimensional problems, examining the dimension scaling of the initialization (Algorithm~\ref{alg:init}) and uncertainty sampling phases of the algorithm. Consider a function $f: [-1,1]^d \rightarrow \mathbb{R}$ with $x := (x_1, \ldots, x_d) \in \mathbb{R}^d$:
\begin{equation}
f(x) = \left \{ \begin{array}{c c}
x^2 + 10 & \quad \text{if } x_d > \sum_{i=1}^{d-1} x_i^3 \\
x^2 - 10 & \quad \text{otherwise}
\end{array}
\right.
\end{equation}
This function is piecewise quadratic with a cubic separating surface. The SVM again employs a Gaussian kernel, and uncertainty sampling adds $N_{add}=10$ points at a time. We vary the parameter dimension $d$ from 2 to 10, and use 10000 points uniformly sampled on the domain of $f$ to evaluate the accuracy of discontinuity localization. For each value of $d$, we run the algorithm until 99\% of these points are classified correctly and plot the total number of function evaluations thus invoked. Results are shown in Figure~\ref{fig:scalePA}. 

Figure~\ref{fig:scalePA} actually shows the results of three experiments, varying the computational effort devoted to the initialization phase of the algorithm. In the first experiment, polynomial annihilation is performed with an edge tolerance $\delta=0.25$ and $N_E=\infty$; in other words, we run Algorithm~\ref{alg:init} until no more refinement is possible, according to the edge tolerance. This is referred to as performing polynomial annihilation ``to completion.'' In the second and third experiments, we set the number of edge points $N_E$ to 20 and 10, respectively. In all three cases, we then proceed with the SVM classification/uncertainty sampling phase of the algorithm until 99\% accuracy is achieved. Solid lines show the total number of function evaluations in each case, while dashed lines show the number of function evaluations performed in the initialization phase only.

The results demonstrate, first of all, that the number of function evaluations required to perform PA to completion grows exponentially with dimension. This effort completely dominates the total computational cost of the algorithm for $d > 3$. But these results also show that performing PA to completion is not necessary. Stopping at $N_E=10$ or $N_E=20$ still allows an accurate characterization of the separating surface to be constructed via uncertainty sampling and the associated labeling. In fact, increasing the number of edge points from 10 to 20 does not really affect the number of training samples subsequently required to achieve 99\% accuracy. One reason this may be the case is that the discontinuity size is fairly consistent across the domain, and therefore once its magnitude is known, additional edge points are unhelpful. As $d$ increases, the efficiency of the algorithm with finite $N_E$ appears to be several orders of magnitude improved over the scenario in which we perform PA until no further refinement is possible.

\subsection{Genetic toggle switch}

A differential-algebraic model of a genetic circuit implemented in \textit{E.\ coli} plasmids has been proposed in \cite{Gardner2000}. This model has been frequently used in the computational science literature as a testbed for uncertainty propagation~\cite{Xiu2007eff}, parameter inference~\cite{MarzoukXiu2009,Moselhy2012}, and discontinuity detection~\cite{Jakeman2011}. The last two studies are concerned with the fact that the system can exhibit a bifurcation with respect to its parameters. The DAE model is as follows:
\begin{eqnarray}
    \frac{du}{dt} &=& \frac{\alpha_1}{1+\nu^{\beta}} - u  \nonumber \\
    \frac{dv}{dt} &=& \frac{\alpha_2}{1+w^{\gamma}} - v  \nonumber \\
    w &=& \frac{u}{(1+\mathrm{[IPTG]}/K)^\eta} \,  .
\label{e:toggle}
\end{eqnarray}
The states $u$ and $v$ essentially represent the expression levels of two different genes. The meanings of the parameters $\alpha_1$, $\alpha _2$, $\beta$, $\gamma$, $\eta$, $K$, and [IPTG] are described in \cite{MarzoukXiu2009} and \cite{Gardner2000}, but are not particularly important here. Following \cite{Jakeman2011} and \cite{Archibald2009}, we fix $\mathrm{[IPTG]} = 4.0 \times 10^{-5}$, $\gamma = 1$, and $\beta = 2.5$.  We consider variation in four parameters, $Z := (\alpha_1, \alpha_2, \eta, K)$. We let $Z$ vary uniformly within a hypercube around the nominal value $Z_0 := (156.25, 15.6, 2.0015, 2.9618\times10^{-5})$, with a range of $\pm 10\%$ in every direction. 
The output of interest is the steady-state value $v(t = \infty)$. 

We apply the discontinuity detection algorithm to this output, with parameters $\delta=\mathtt{tol}=0.25$, $N_E = 15$, $\delta_t=2.0$, and $\epsilon=0.01.$ The PA parameter choices $\delta$, $\mathtt{tol}$, and $N_E$ are all targeted towards achieving a small number of function evaluations during PA. In this model, it is known that the function values exhibit small variability compared to the jump size. For this reason we only need to learn the function values in each class at a few locations in the parameter domain. Once we obtain this information, we have essentially learned the jump size over the entire parameter domain. We can also use a fairly large $\delta_t$, in this case encompassing a majority of the parameter domain. These choices demonstrate the flexibility of the algorithm, in that the parameter choices can reflect prior knowledge when it is available. Finally, the algorithm is fairly insensitive to $\epsilon$ because we will employ a stop-short mechanism for termination.

Table~\ref{tab:toggle} shows the number of model evaluations (i.e., integrations of (\ref{e:toggle})) required to localize the discontinuity to within a 1\% classification error and a 0.1\% classification error, measured with 5000 random samples from the parameter domain. Since our algorithm involves random sampling, we report the average and standard deviation of this value over 100 independent runs. For comparison, we also report the number of model evaluations used by the edge tracking scheme of \cite{Jakeman2011} and adaptive refinement scheme of \cite{Archibald2009} for exactly the same problem. The performance of the new discontinuity localization algorithm is much improved over these previous techniques. We attribute the improvement to the fact that the separating surface in this example can be well described using a hyperplane, as in \cite{Archibald2009}. The SVM classifer needs very few points to approximate a hyperplane or small perturbations thereof.

\begin{table}[H]
\begin{center}
\begin{tabular}{|l|c|c|c|}
\hline
& Learning & Edge Tracking \cite{Jakeman2011}  & Adaptive Refinement\cite{Archibald2009} \\
\hline
 model evals for 1\% error & 127 $\pm$ 2 & 31,379& 91,250 \\
 model evals for 0.1\% error & 257 $\pm$ 22 & -- & -- \\
\hline
\end{tabular} 
\end{center}
\vspace{6pt}
\caption{Performance comparison of different discontinuity localization schemes on the genetic toggle switch example.}
\label{tab:toggle}
\end{table}

\subsection{Discontinuity in subspace of full domain}

Finally, we demonstrate the performance of the discontinuity detection scheme on a problem wherein the discontinuity occurs only along a subset of the input coordinates. In other words, the problem contains a separating surface that is aligned with a complementary subset of the coordinate directions. In particular, we consider a function $f: [-1,1]^{20} \rightarrow \mathbb{R}$ containing a discontinuity across a 2-sphere that is ``extruded'' through a 20-dimensional ambient space: 
\begin{equation*}
f(x) = \left\{
\begin{array}{rl}
1 & \mathrm{if}\quad \displaystyle \sum_{i=1}^{3} x_i^2 < r^2 \\
-1 & \mathrm{else} .
\end{array} \right.
\end{equation*}
Here $x \in [-1,1]^{20}$ and $x_i$ is the $i$th component of $x$. The radius $r = 0.125$. This test case was first proposed in \cite{Jakeman2011}. 
Note that the separating surface here is still a 19-dimensional manifold. 

In this example, we evaluate the classification error at 1000 random points uniformly sampled over the 20-dimensional subregion located within a distance 0.125 of the separating surface. If the 1000 uniformly random points instead covered the full 20-dimensional hypercube, the classification error would be excessively low; focusing on the region near the discontinuity, on the other hand, provides a more stringent test of our approximation scheme. To compare our algorithm's performance with that of the edge tracking scheme in \cite{Jakeman2011}, uncertainty sampling iterations are continued until we achieve 93\% classification error in the subregion. We use the same algorithm input parameters as in the previous example, with the exception of seeking only one edge point in each dimension, starting with an initial point set $\mathcal{M}_0$ that contains only the origin. The edge tracking results indicated that $\mathcal{O}(10^4)$ function evaluations were required to achieve 93\% accuracy. The new discontinuity detection approach, on the other hand, requires 275 function evaluations function evaluations in the initialization phase and approximately 200 function evaluations during uncertainty sampling. 
Again, this improvement reflects the fact that $f$ contains a very regular discontinuity shape, which can be approximated by the kernel SVM quite efficiently.

%% file: conc-ack.tex
\section{Conclusions}

This paper has developed an efficient and flexible discontinuity localization algorithm for the outputs of parameter-dependent computational simulations. The algorithm progressively refines a functional approximation of the surface across which the discontinuity occurs. The approach is unstructured; after an initialization phase employing polynomial annihilation, it relies on guided random sampling and thus avoids constructing a dense structured grid of model evaluations, either globally or in the vicinity of the discontinuity. The separating surface is represented by a kernel SVM classifier; this representation is quite flexible, and is able to capture a wide range of discontinuity geometries over a range of dimensions. We demonstrate the approach on several model functions and benchmark ODE and PDE systems. Compared to previous approaches, it requires significantly fewer model evaluations to achieve a given level of accuracy.

The advantage of this algorithm is greatest when the complexity of the separating surface is low, so that it can be well approximated with few points. But the nonparametric representation employed by the SVM allows it to approximate surfaces ranging from linear (i.e., hyperplanes) to very complex (i.e., disconnected). By contrast, other unsupervised discontinuity localization schemes assume that the geometry of the separating surface necessitates some form of edge tracking (in a sense allowing maximum complexity, up to a point-spacing tolerance $\delta$), or assume a particular parameterization of the surface. In return for the present flexibility, one must make some assumptions about how quickly function values near the separating surface vary relative to the local jump size. When little is known, one can make conservative choices of the variation radius $\delta_t$ and the edge tolerance $\delta$, and the method in a sense reverts to a Monte Carlo edge tracking scheme.

While we have demonstrated the effectiveness of the algorithm on a wide variety of problems, future refinements can extend it to simulations whose input domains must be divided into more than two classes, perhaps resulting from several disconnected output regimes, and to simulations that exhibit discontinuities in their derivatives. Tackling these issues should not fundamentally change the present methodology; indeed, one could employ multi-class SVM classifiers with an appropriate labeling scheme. Finally, we emphasize that the localization of discontinuities is but one step towards the development of efficient surrogates for computational simulations. Future work will couple the methodology presented here with function approximation techniques to create a unified framework for the construction of piecewise smooth surrogate models.

\subsection*{Acknowledgments}

The authors gratefully acknowledge BP for funding this research.